%% file: main.tex
\documentclass{article}

\usepackage{microtype}
\usepackage{graphicx}
\usepackage{subcaption}
\usepackage{booktabs} %
\usepackage[utf8]{inputenc}
\usepackage[T1]{fontenc}    %
\usepackage{wrapfig}
\usepackage{tcolorbox}
\usepackage{lipsum}  %

\usepackage{arxiv}
\setcitestyle{authoryear,open={(},close={)}}
\usepackage[subtle]{savetrees}
\usepackage{titling}

\def\shownotes{1}  %
\ifnum\shownotes=1
\newcommand{\authnote}[2]{[#1: #2]}
\else
\newcommand{\authnote}[2]{}
\fi
\newcommand{\subtitle}[1]{%
  \posttitle{%
    \par\end{center}
    \begin{center}\large#1\end{center}
    \vskip0.5em}%
}

\def\arxiv{1}
\ifnum\arxiv=1

\else

\fi

\newcommand{\name}[0]{WORKBank\xspace}

\begin{document}
\bibliographystyle{plainnat}

\title{Future of Work with AI Agents: \\ \vspace{8pt} \Large Auditing Automation and Augmentation Potential across the U.S. Workforce}

\author{
     \large{Yijia Shao\textsuperscript{*}, Humishka Zope\textsuperscript{*}, Yucheng Jiang, Jiaxin Pei, David Nguyen}, \\\large{Erik Brynjolfsson, Diyi Yang} \\
        \large{Stanford University}\\
        {\texttt{\{shaoyj, diyiy\}@cs.stanford.edu}}\\
    Data \& Code: \href{https://futureofwork.saltlab.stanford.edu/}{\url{futureofwork.saltlab.stanford.edu}}
}

\date{}
\def\thefootnote{*}\footnotetext{Equal Contribution}\def\thefootnote{\arabic{footnote}}

\maketitle

\doparttoc %
\faketableofcontents %

\begin{abstract}
\input{0-Abstract}

\end{abstract}

\section{Introduction}
\input{1-Introduction}

\section{Auditing Framework}
\input{3-Methods}

\section{Results}
\input{4-Results}

\section{Related Work}
\input{2-Related_Work}

\section{Conclusion}
\input{8-Discussion}

\section*{Acknowledgements}
\input{9999-Acknowledgement}

\bibliography{ref}

\newpage
\appendix

\addcontentsline{toc}{section}{Appendix} %
\part{Appendix} %
\parttoc %
\clearpage
\newpage
\input{999-Appendix}

\end{document}

%% file: 0-Abstract.tex
The rapid rise of compound AI systems (\aka, AI agents) is reshaping the labor market, raising concerns about job displacement, diminished human agency, and overreliance on automation. Yet, we lack a systematic understanding of the evolving landscape. In this paper, we address this gap by introducing a novel auditing framework to assess which occupational tasks workers want AI agents to automate or augment, and how those desires align with the current technological capabilities. Our framework features an audio-enhanced mini-interview to capture nuanced worker desires and introduces the Human Agency Scale (HAS) as a shared language to quantify the preferred level of human involvement. Using this framework, we construct the \name database, building on the U.S. Department of Labor's O*NET database, to capture preferences from 1,500 domain workers and capability assessments from AI experts across over 844 tasks spanning 104 occupations. Jointly considering the desire and technological capability divides tasks in \name into four zones: Automation ``Green Light'' Zone, Automation ``Red Light'' Zone, R\&D Opportunity Zone, Low Priority Zone. This highlights critical mismatches and opportunities for AI agent development. Moving beyond a simple automate-or-not dichotomy, our results reveal diverse HAS profiles across occupations, reflecting heterogeneous expectations for human involvement. Moreover, our study offers early signals of how AI agent integration may reshape the core human competencies, shifting from information-focused skills to interpersonal ones. These findings underscore the importance of aligning AI agent development with human desires and preparing workers for evolving workplace dynamics.

%% file: 1-Introduction.tex
Rapid advances in foundation models, such as large language models (LLMs), have catalyzed growing interest in AI agents: goal-directed systems equipped with tool access and multi-step execution capabilities. Unlike standalone models, these agents can perform complex workflows and are increasingly positioned to take on roles across a broad range of professional domains~\citep{yang2024swe,openhands,shao-etal-2024-assisting,jiang-etal-2024-unknown,yao2024taubench}. Their integration into occupational settings is already beginning to shape the labor market~\citep{hoffmann2024generative,demirci2025ai}. 
For example, research indicates that around 80\% of U.S. workers may see LLMs affect at least 10\% of their tasks, with 19\% potentially seeing over half impacted~\citep{eloundou2023gptsgptsearlylook}. Usage data from Anthropic indicates that in early 2025, at least some workers in 36\% of occupations \textit{already} were using AI for at least 25\% of their tasks~\citep{handa2025economic}.

While AI adoption in the workplace has shown promise in boosting productivity, it also raises concerns about job displacement, reduced human agency, and overreliance on automation~\citep{hazra2025ai}. Despite this critical impact, we lack a systematic and grounded understanding of the evolving landscape. From a coverage perspective, prior research often focuses on a few domains like software engineering~\citep{hoffmann2024generative} and customer support~\citep{brynjolfsson2025generative}. This narrow scope limits our comprehension of the real-world complexity of diverse human jobs and the varied nature of open-ended tasks. From a stakeholder perspective, existing studies often emphasize the interests of capital by focusing on a few tasks that tend to be more profitable such as coding without adequately considering worker values~\citep{eisfeldt2023labor}. Furthermore, current approaches often rely on analyzing existing usage data, such as how people use chatbots for work~\citep{hazra2025ai, zhao2024wildchat}, which cannot provide a forward-looking assessment of AI potential across the broader workforce.

To address these gaps, we propose a principled, survey-based framework to investigate which occupational tasks workers want AI agents to automate or augment. %
We look at the entire workforce that could be impacted by digital AI agents by sourcing occupational tasks from the U.S. Department of Labor's O*NET database. Compared to occupation-level studies, task-level auditing allows us to better capture the nuanced, open-ended, and contextual nature of real-world work. Our auditing framework takes a worker-centric approach by soliciting first-hand insights from domain workers actively performing the tasks. To guide domain workers in providing well-calibrated responses, we empower them to share their experiences and articulate their reasoning through an audio-enhanced survey system. 
Crucially, our framework expands beyond the binary view of automation. We propose the Human Agency Scale (\ie, H1-H5), which complements the SAE L0-L5 automation levels~\citep{sae2014taxonomy} by quantifying the degree of human involvement required for occupational task completion and quality. This new scale centers human agency---a crucial factor for responsible AI agent adoption~\citep{fanni2023enhancing}---and provides a shared language to capture the spectrum between automation and augmentation.%

To ground workers' perspectives in technical reality, we further gather complementary assessments from AI experts with experience in agent research and development (R\&D). This dual approach reveals how workers and experts perceive AI agents' capabilities and risks at work.  Based on data collected through January 2025 to May 2025, we construct the AI Agent \textbf{W}orker \textbf{O}utlook \& \textbf{R}eadiness \textbf{K}nowledge Bank (\name). This database currently consists of responses from 1,500 workers across 104 occupations and annotations from 52 AI experts, covering 844 occupational tasks. It is designed to be easily extensible to more tasks and to reflect evolving technological capabilities and worker preferences. To our knowledge, this is the first large-scale audit of AI agent capabilities and worker preferences. 

Our work contributes four sets of findings: 
\begin{enumerate}
    \item {\textbf{Domain workers want automation for low-value and repetitive tasks (\reffig{Fig:automation_desire}).} For 46.1\% of tasks, workers express positive attitudes toward AI agent automation, even after reflecting on potential job loss concerns and work enjoyment. The primary motivation for automation is freeing up time for high-value work, though trends vary significantly by sector.} 
    \item {\textbf{We visualize the desire-capability landscape of AI agents at work, and find critical mismatches (\reffig{Fig:automation_readiness}).} The worker desire and technological capability divide the landscape into four zones: Automation ``Green Light'' Zone (high desire and capability), Automation ``Red Light'' Zone (high capability but low desire), R\&D Opportunity Zone (high desire but currently low capability), and Low Priority Zone (low desire and low capability). Notably, 41.0\% of Y Combinator company-task mappings are concentrated in the Low Priority Zone and Automation ``Red Light'' Zone. Current investments mainly center around software development and business analysis, leaving many promising tasks within the ``Green Light'' Zone and Opportunity Zone under-addressed.}
    \item {\textbf{The Human Agency Scale provides a shared language to audit AI use at work and reveals distinct patterns across occupations (\reffig{Fig:has_overview}).} 45.2\% of occupations have H3 (equal partnership) as the dominant worker-desired level, underscoring the potential for human-agent collaboration. However, workers generally prefer higher levels of human agency, %
    potentially foreshadowing friction as AI capabilities advance.}
    \item {\textbf{Key human skills are shifting from information processing to interpersonal competence (\reffig{Fig:skill_shift}).} By mapping tasks to core skills and comparing their associated wages and required human agency, we find that traditionally high-wage skills like analyzing information are becoming less emphasized, while interpersonal and organizational skills are gaining more importance. Additionally, there is a trend toward requiring broader skill sets from individuals. These patterns offer early signals of how AI agent integration may reshape core competencies.}
\end{enumerate}

%% file: 3-Methods.tex
\label{sec:framework}

\begin{figure*}[t]
    \centering
    \resizebox{\textwidth}{!}{%
    \includegraphics{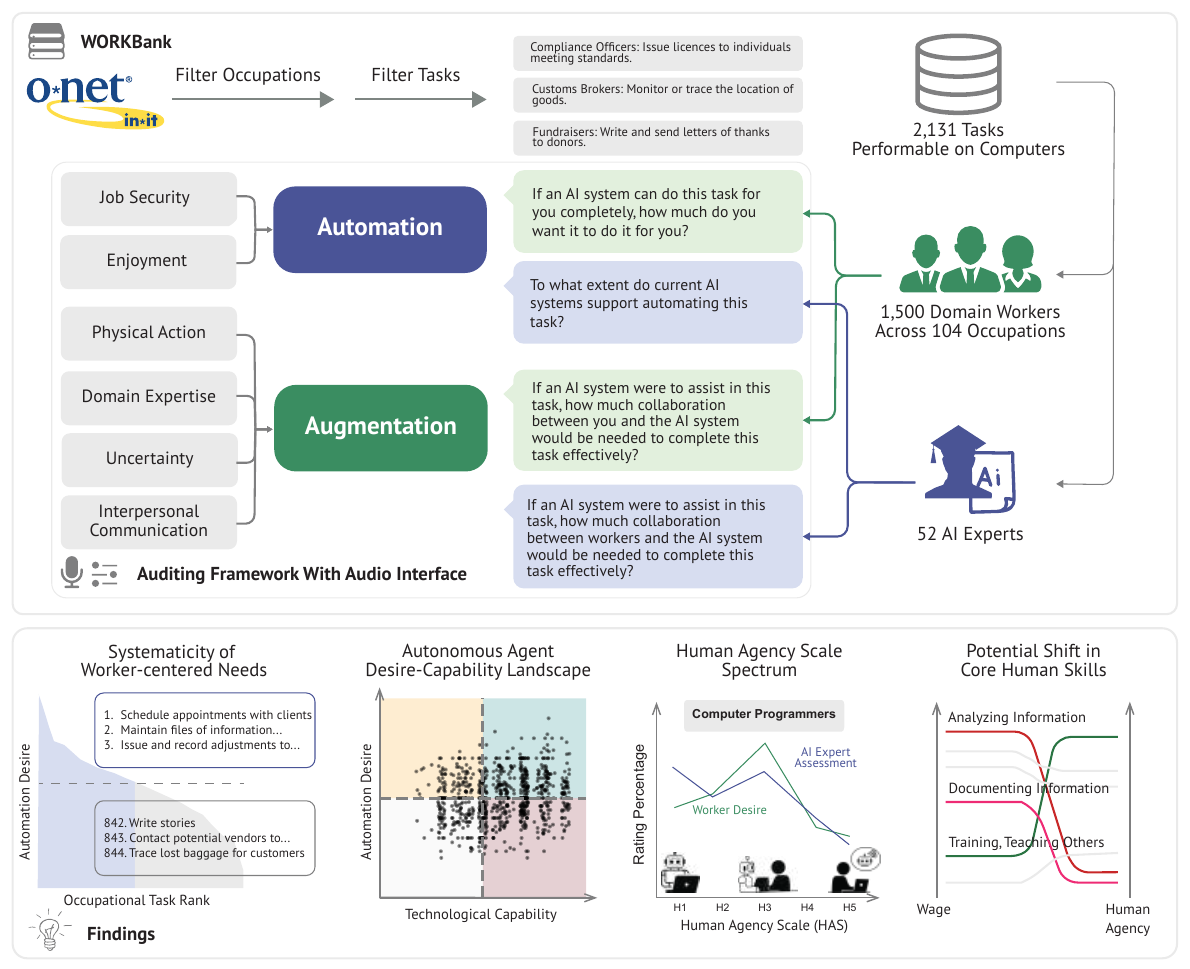}
    }
    \caption{\textbf{Overview of the auditing framework and key insights.} The framework captures dual perspectives on automation and augmentation by eliciting both worker desires and expert assessments of technological capabilities. It guides participant reasoning through structured prompts and an audio-enhanced interface. We instantiate this framework to build the \name database, enabling a data-driven analysis of worker-centered needs, the desire–capability landscape, the Human Agency Scale (HAS) spectrum, and implications for core human skills.
    }
    \label{Fig:audit_framework}
\end{figure*}

\begin{figure*}[t]
    \centering
    \resizebox{\textwidth}{!}{%
    \includegraphics{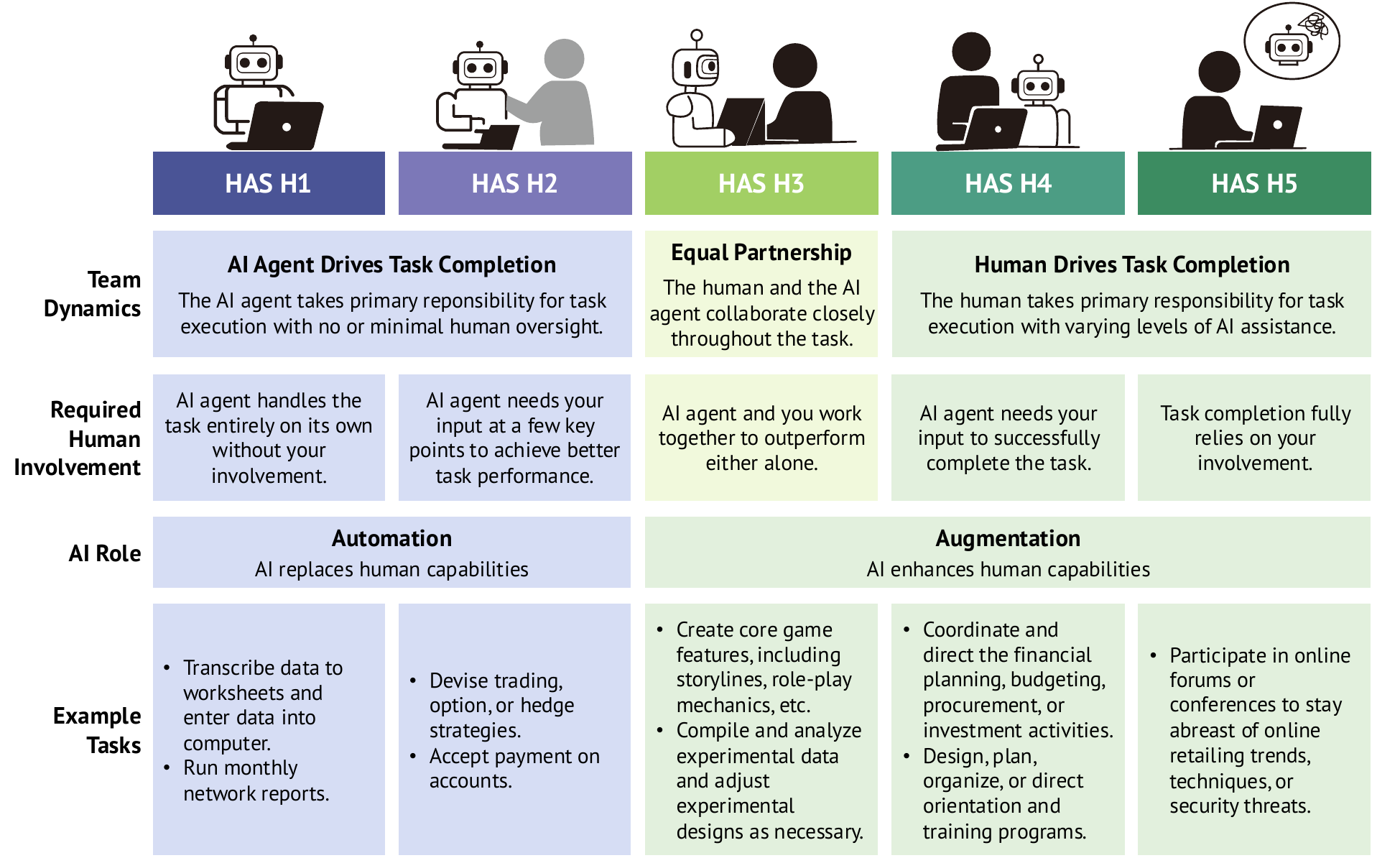}
    }
    \caption{\textbf{Levels of Human Agency Scale (HAS).} We introduce the Human Agency Scale (\ie, H1-H5) to quantify  the team dynamics and degree of human involvement required. HAS provides a shared language to quantify automation vs. augmentation, complementing the traditionally ``AI-first'' perspective used in defining levels of automation. Importantly, higher HAS levels are not inherently better---different levels suit different AI roles.
    }
    \label{Fig:has}
\end{figure*}

To investigate how AI agents may integrate into professional work, we develop a task-level, survey-based auditing framework that captures both worker preferences and technological feasibility across the automation–augmentation spectrum. We begin by outlining a few key design principles of our framework before examining each one in detail.

\subsection{Defining Audit Granularity and Scope}
Our framework focuses on \textit{complex, multi-step tasks} associated with specific occupations (\eg, ``Marketing Managers: Compile lists describing product or service offerings''), sourced from the O*NET database. These tasks, unlike isolated, low-level activities (\eg, ``track goods or materials'' or ``translate information''), reflect actual job responsibilities and the kinds of workflows AI agents are poised to impact. Moreover, compared with occupation-level analysis, conducting the audit at the task level enables a more nuanced understanding, as tasks within the same profession can vary significantly and are often highly contextualized.

We scope our audit to computer-compatible tasks, recognizing their susceptibility to \textit{foundation model-powered AI agents}. Drawing from historical and recent literature on agent autonomy~\citep{reason:RusNor95a, wooldridge1995intelligent, castelfranchi1994guarantees}, planning capabilities, and tool use~\citep{mitchell2025fully}, we define AI agents (excluding physical robots) as: \textit{``A system or program capable of autonomously performing tasks on behalf of a user or another system by designing its workflow and utilizing available software tools, without the ability to perform physical actions.''}

\subsection{Emphasizing the Spectrum of Automation and Augmentation}
Traditional technology impact studies often ask: \textit{To what degree can this task be automated?} Besides this view of automation, we consider the view of augmentation---where technology complements and enhances human capabilities~\citep{autor2015there}, as this new wave of technology holds significant potential to augment human workers through human-agent collaboration, enhancing both productivity and work quality. %
While augmentation has been discussed in prior work~\citep{brynjolfsson2022turing,handa2025economic}, there is no established framework for quantifying automation vs. augmentation. 
To fill this gap and provide a shared language, we introduce the Human Agency Scale (HAS) (\reffig{Fig:has}), a five-level scale from H1 (no human involvement) to H5 (human involvement essential):
\begin{itemize}\setlength\itemsep{0em}
    \item {H1: AI agent handles the task entirely on its own.}
    \item {H2: AI agent needs minimal human input for optimal performance.}
    \item {H3: AI agent and human form equal partnership, outperforming either alone.}
    \item {H4: AI agent requires human input to successfully complete the task.}
    \item {H5: AI agent cannot function without continuous human involvement.}
\end{itemize}

Unlike SAE driving automation levels~\citep{sae2014taxonomy} that adopt an ``AI-first'' perspective, HAS provides a human-centered lens for assessing both task properties and appropriate agent development approaches. Importantly, higher HAS levels are not inherently better---different levels suit different AI roles. Tasks at H1-H2 favor automation approaches, while H3-H5 tasks benefit from augmentation strategies. Understanding the ideal level of human involvement is essential both for workers seeking to adapt their skills and for developers aiming to build context-appropriate AI agents. For instance, fully autonomous agents shall be developed for H1 scenarios, while those agents for H3 scenarios must support meaningful coordination and communication with human collaborators~\citep{shao2024collaborative}.

This five-level human agency scale (H1-H5) helps categorize tasks where AI is more suitable for automation (H1-H2) versus augmentation (H3-H5), where human agency remains critical.

\subsection{Constructing A Worker-Centric Auditing Framework}
\label{sec:worker_value}
Our auditing framework centers on the needs of workers. To support domain workers in providing well-calibrated feedback, we enable them to share their experiences and explain their thought processes using an audio-supported survey system. Concretely, for each task $t$, we first collect worker ratings on \textit{automation desire $A_w(t)$} and \textit{desired HAS level $H_w(t)$} using a 5-point Likert scale.

\input{prompts/automation_desire_likert_question}

\input{prompts/desired_has_level_question}

To support thoughtful ratings, we also scaffold worker responses through three key designs (survey details in Appendix~\ref{appendix:survey_details}):
\begin{itemize}
    \item {\textbf{Audio-enhanced Reflection}: The survey begins with an audio-enhanced mini-interview exploring participants' occupational work and AI perspectives. This spoken format enables more natural reflection and helps workers more efficiently contextualize their ratings within their actual work experience, compared to slowly typing their experiences.}
    \item {\textbf{Quality Control via Task Familiarity Filtering}: Workers receive only occupation-relevant tasks and must confirm task familiarity before rating, ensuring assessments are grounded in their real experience rather than speculation.}
    \item{\textbf{Guided Consideration}: Before rating both automation desire $A_w(t)$ and desired HAS level $H_w(t)$, participants consider factors identified in prior literature---enjoyment and job security concerns for automation desire~\citep{armstrong2024automationworkersperspective, GODOLLEI2023100342}, and task characteristics like physical actions, domain expertise requirements, uncertainty, and interpersonal elements for HAS preferences~\citep{shah2024agents, parasuraman2000designing, frank2019toward}.}
\end{itemize}

\subsection{Ensuring Dual Perspectives from Both Domain Workers and AI Experts}
While worker perspectives provide invaluable insights into the social demand and acceptance of AI agents, they represent only one side of the integration equation. Domain workers, despite their deep task expertise, may have limited exposure to current AI capabilities and constraints.
Thus, we complement workers' perspectives with expert assessments of \textit{current automation capability $A_e(t)$} and \textit{feasible HAS level $H_e(t)$} from AI researchers and practitioners. %
This dual perspective reveals the readiness for AI agent integration and allows us to identify alignment or gaps between worker desires and technological feasibility.

Concretely, these experts assess $A_e(t)$ and $H_e(t)$ using the same rubrics, drawing on their understanding of existing systems' strengths and limitations. Contrasting $A_w(t)$, $H_w(t)$ and $A_e(t)$, $H_e(t)$ enables us to understand what require future breakthroughs, identify alignments and misalignments between worker preferences and technological development, and inform development priorities.

\subsection{Instantiating the Audit Framework to Derive \name}
\label{sec:dataset}

\begin{figure*}[th]
    \centering
    \resizebox{\textwidth}{!}{%
    \includegraphics{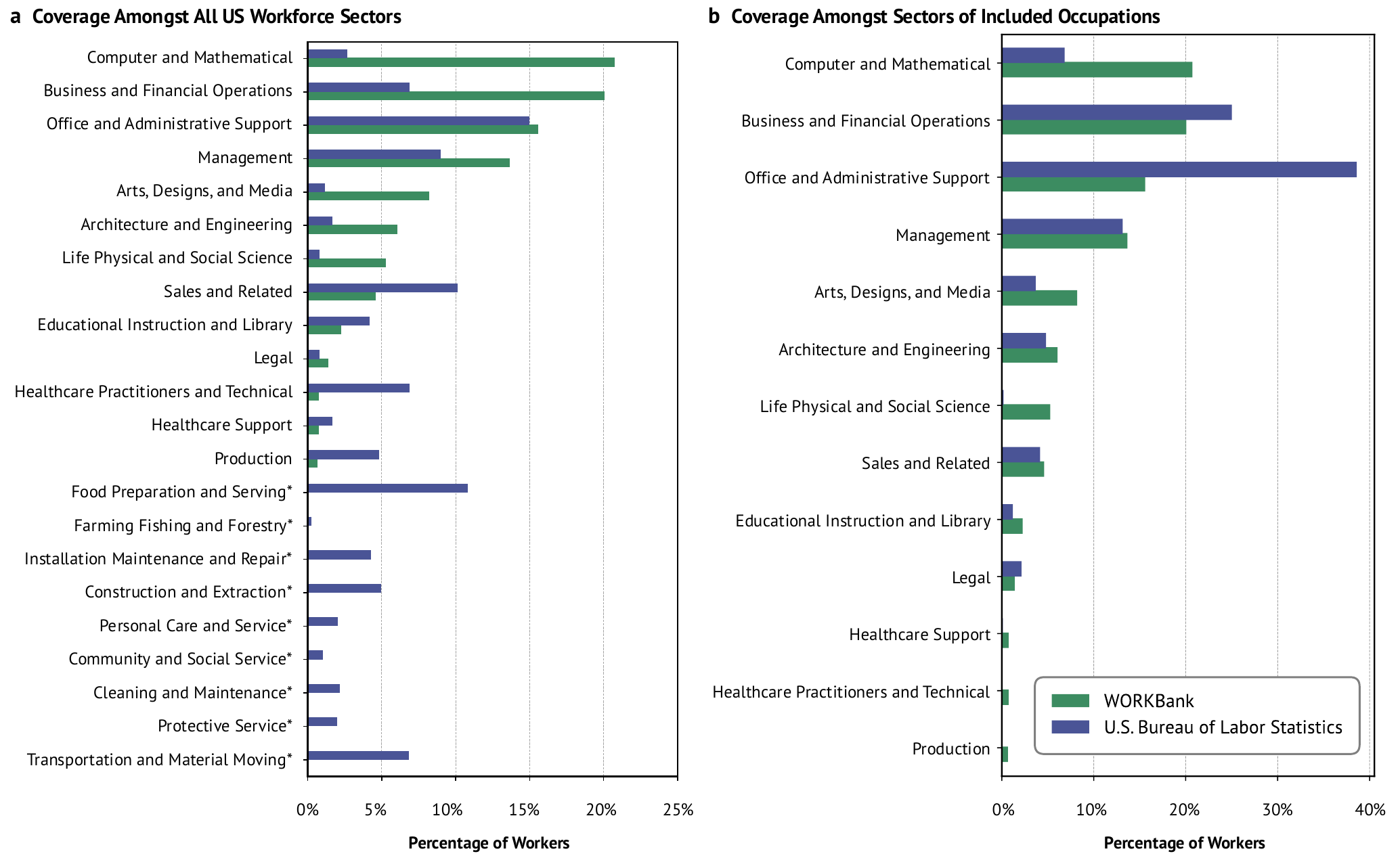}
    }
    \caption{\textbf{Sector-level distribution of workers in the \name database compared to U.S. workforce statistics from the Bureau of Labor Statistics.}
    \textbf{a}, Comparison between \name worker distribution and the U.S. workforce employment statistics across all sectors (sectors not included in \name marked with an asterisk).
    \textbf{b}, Comparison between \name worker distribution and the U.S. workforce employment statistics limited to the 104 occupations included in our database.}
    \label{Fig:sector_dist}
\vspace{-2em}
\end{figure*}

Taking into account these aforementioned design principles together, we then apply our auditing framework to develop \textbf{W}orker \textbf{O}utlook \& \textbf{R}eadiness \textbf{K}nowledge Bank (\name). Concretely, we source computer-compatible tasks performed at least monthly from the U.S. Department of Labor's O*NET Database (details in Appendix~\ref{appendix:task_source}). These tasks reflect complex, multi-step workflows central to our focus. For example, the task ``Credit Analysts: Analyze credit data and financial statements to determine the degree of risk involved in extending credit or lending money'' entails data analysis, risk evaluation, and decision-making. After filtering, 2,131 tasks across 287 occupations remain.

We developed an IRB-approved, self-hosted survey interface and distributed it through crowdsourcing platforms and targeted LinkedIn outreach. Between January and May 2025, recruitment through Prolific, Upwork, and LinkedIn yielded 1,678 participants who provided 7,016 task ratings. After filtering for adequate representation ($\geq10$ participants per occupation), we obtained assessments from 1,500 individuals across 104 occupations and calculated average worker ratings $A_w(t)$ and $H_w(t)$ for each task. We evaluate the representativeness of \name by comparing its sector-level distribution with U.S. workforce data from the Bureau of Labor Statistics (Appendix~\ref{appendix:workforce_coverage}). As shown in~\reffig{Fig:sector_dist}, the comparisons suggest that our database captures a broad and representative cross-section of the U.S. workforce at the sector level.

For expert assessments, we recruited 52 AI experts---PhD researchers and industry practitioners with experience in AI agent R\&D. Each task was independently assessed by at least two experts, with additional reviews ensuring rating consistency (standard deviation $\leq1$). Inter‐annotator agreement, measured by Krippendorff’s $\alpha$, was 0.539 for $A_e(t)$ and 0.511 for $H_e(t)$ (see robustness analysis details in Appendix~\ref{appendix:robustness}).

Combining these complementary data sources, we construct \textbf{W}orker \textbf{O}utlook \& \textbf{R}eadiness \textbf{K}nowledge Bank (\name), the first database to capture both worker desires and AI agents' technological capabilities for occupational tasks.

%% file: prompts/automation_desire_likert_question.tex
{
\begin{tcolorbox}[colframe=black, colback={rgb,255:red,250;green,250;blue,250}, colbacktitle=gray!50!black, fontupper=\small, fonttitle=\bfseries\color{white}, title=Likert Question for Collecting Automation Desire]
If an AI system can do this task for you completely, how much do you want it to do it for you?\\
1: Not at all; 2: Slightly; 3: Moderately; 4: A lot; 5: Entirely

\end{tcolorbox}

\captionsetup{type=figure}
\label{prompt:automation_desire_question}
}

%% file: prompts/desired_has_level_question.tex
{
\begin{tcolorbox}[colframe=black, colback={rgb,255:red,250;green,250;blue,250}, colbacktitle=gray!50!black, fontupper=\small, fonttitle=\bfseries\color{white}, title=Likert Question for Collecting Desired HAS Level]
If an AI system were to assist in this task, how much collaboration between you and the AI system would be needed to complete this task effectively?\\
1: No Collaboration Needed (Human Agency Scale H1);\\
2: Limited Collaboration Needed (Human Agency Scale H2);\\
3: Moderate Collaboration Needed (Human Agency Scale H3);\\
4: Considerable Collaboration Needed (Human Agency Scale H4);\\
5: Essential Collaboration Needed (Human Agency Scale H5)

\end{tcolorbox}

\captionsetup{type=figure}
\label{prompt:has_desire_question}
}

%% file: 4-Results.tex
\label{sec:results}

Leveraging the rich data within \name, we examine where workers most desire automation by AI agents and where they resist it, whether current AI capabilities and R\&D align with these preferences, what opportunities exist for AI to augment rather than replace human labor, and how the presence of AI agents might reshape the demand for human skills.

\subsection{Worker-centered Views on Occupational Task Automation}
\label{sec:automation_desire}

\begin{figure*}[th]
    \centering
    \resizebox{\textwidth}{!}{%
    \includegraphics{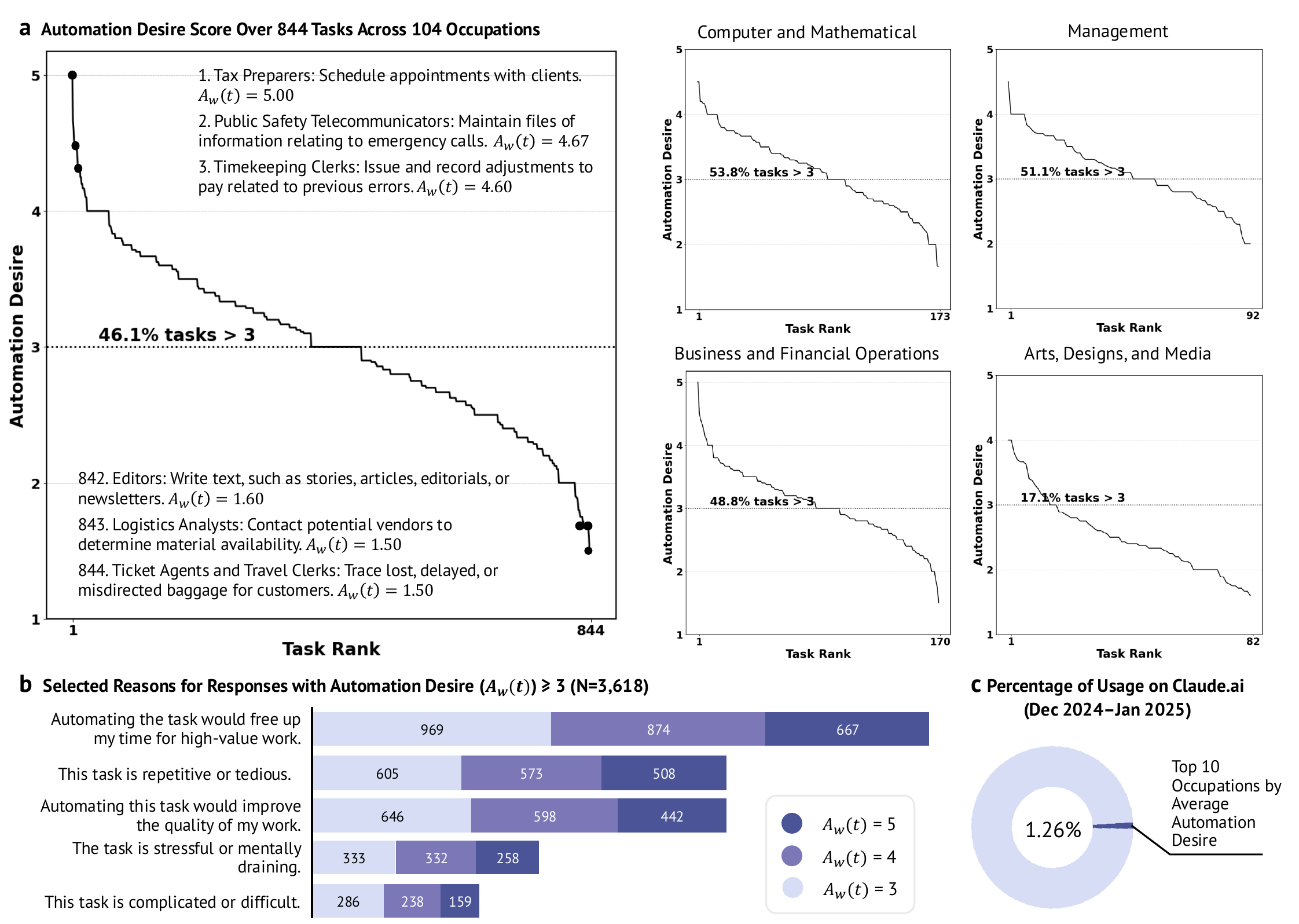}
    }
    \caption{\textbf{First-hand data from domain workers reveals positive attitudes towards AI agent automation on certain occupational tasks, particularly due to perceived benefits such as freeing up time for high-value work. However, the sentiment varies notably across sectors.} \textbf{a}, Automation desire scores $A_w(t)$ over 844 occupational tasks, ranked based on \name data, together with sector-specific breakdowns. The distribution indicates a mixed attitude, revealing high diversity of needs and preferences of workers that should be considered in AI agent R\&D. \textbf{b}, Reported reasons for responses with $A_w(t)\geq3$. The most selected reason---``Automating the task would free up my time for high-value work''---accounts for 69.38\% of the responses. \textbf{c}, Comparison with usage data from Claude.ai, a LLM-based chatbot (Dec 2024-Jan 2025, from \citet{handa2025economic}), shows that the top 10 occupations with the highest average automation desire represent only 1.26\% of total usage. This highlights the importance of directly soliciting worker input, as usage data may lag behind actual workplace needs.
    }
    \label{Fig:automation_desire}
\end{figure*}

\paragraph{Where do workers desire AI agent automation?}
We first examine domain workers’ attitudes toward automating their occupational tasks. %
In \reffig{Fig:automation_desire} \textbf{a}, we rank tasks by their average worker automation desire scores $A_w(t)$. We find that for 46.1\% of tasks, workers currently performing them express a positive attitude (\ie, $A_w(t)>3$) toward AI agent automation, even after explicitly considering concerns such as job loss and reduced enjoyment, as guided by our auditing framework.  On the other hand, the distribution indicates a mixed attitude, with 7.11\% tasks receiving $A_w(t)\geq4$ and 6.16\% receiving $A_w(t)\leq2$. 
To better understand these preferences, \reffig{Fig:automation_desire} \textbf{b} aggregates selected reasons given for pro-automation responses ($A_w(t)\geq3$). The most cited motivation---``freeing up time for high-value work''---was selected in 69.38\% of cases. Other common reasons include task repetitiveness (46.6\%), stressfulness (25.5\%), and opportunities for quality improvement (46.6\%). The overall pattern suggests that AI agents could play a supportive role, enabling workers to offload low-value or burdensome tasks, rather than serving as replacements in a zero-sum dynamic.

\paragraph{Does existing LLM usage reflect worker desires?}
Notably, when we compare our findings with usage data from Claude.ai, an LLM-based chatbot used between Dec 2024 and Jan 2025~\citep{handa2025economic}, we find that the top 10 occupations with the highest average automation desire account for only 1.26\% of total usage. This mismatch highlights a disconnect: occupations where workers most desire automation are currently underrepresented in LLM usage. This suggests that existing usage patterns may be skewed toward early adopters or specific job types, rather than reflecting broader demand. Such a gap reinforces the value of our worker-centric audit, which surfaces latent needs that may not yet appear in usage logs.%

\paragraph{Where do workers resist AI agent automation?}
We analyze audio response data using LLM-based topic modeling to identify the primary concerns workers expressed regarding the use of AI agents in their work (see Appendix~\ref{sec:topic_modeling}). Among our survey participants, 28.0\% expressed fears, concerns, or negative sentiment when answering the question ``How do you envision using AI in your daily work?''. Among these workers, the three most prominent concerns identified are: (1) lack of trust in AI systems' accuracy, capability, or reliability (45.0\%), (2) fear of job replacement (23.0\%), and (3) the absence of human qualities or capabilities in AI 
 (16.3\%).

When discussing the absence of human qualities, workers express specific concerns about losing a ``human touch'' in their work, diminishing creative control, and the desire to maintain agency in decision-making. This sentiment echoes our quantitative findings from the breakdown of automation desire scores across sectors (Figure~\ref{Fig:automation_desire} \textbf{a}). In these sector-level breakdowns, the ``Arts, Designs, and Media'' sector stands out, with only 17.1\% of tasks receiving positive desire ratings ($>3$ on a 5-point Likert scale). Audio responses from participants in this sector reveal nuanced opposition to automating content creation, such as: ``\emph{I want it to be used for seamlessly maximizing workflow and, you know, making things less repetitive and tedious and arduous with workflow. No content creation,}'' ``\emph{I would never use AI to like replace artists. I would be more for personal [project management] use, if anything,}'' ``\emph{AI can be a game-changer in data architect workflow, helping to improve efficiency, accuracy and even creativity. But I create my design by myself. For research, I use AI}''.

\begin{figure*}[t]
    \centering
    \resizebox{0.9\textwidth}{!}{%
    \includegraphics{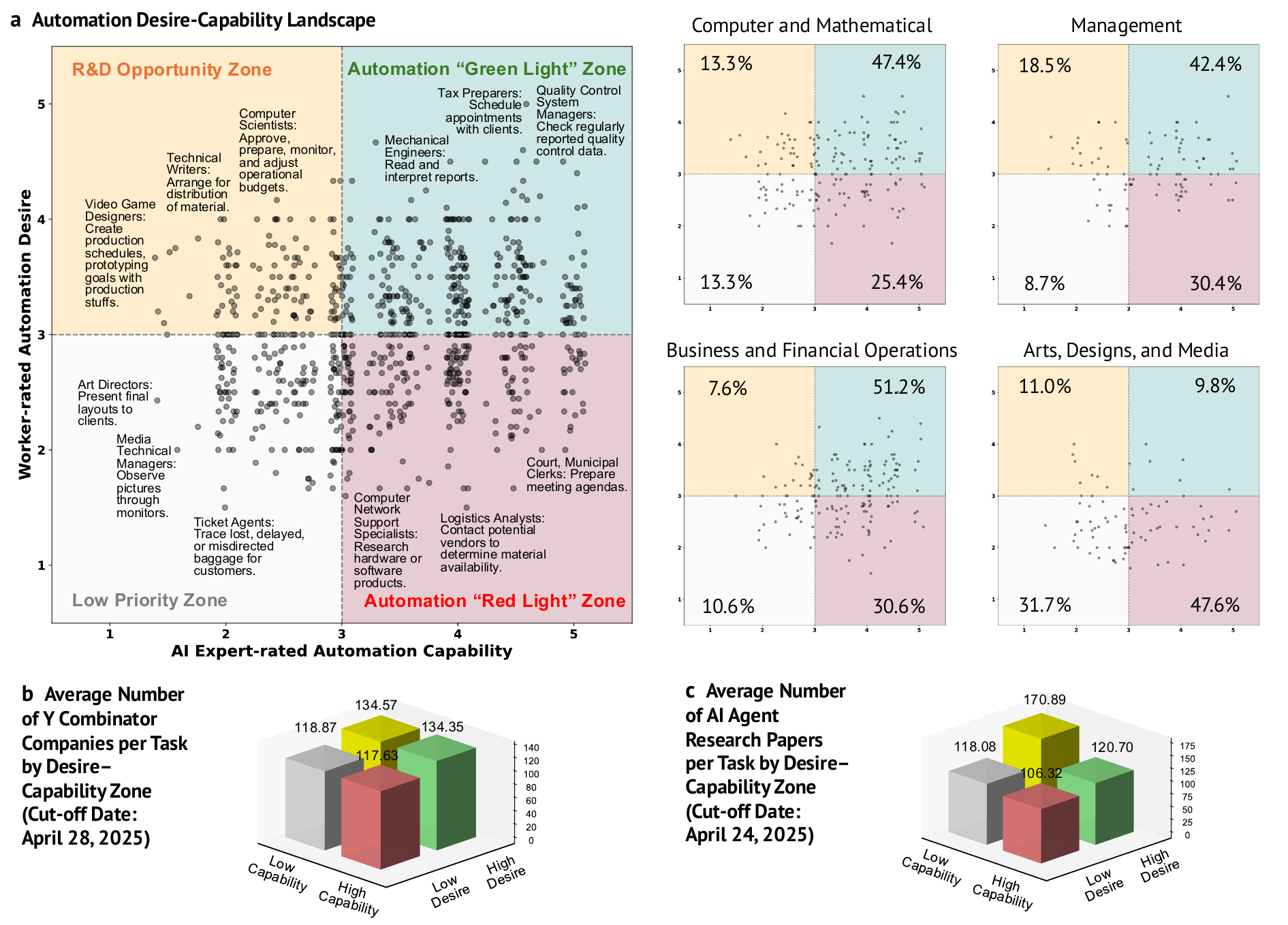}
    }
    \caption{\textbf{Integrating worker and AI expert perspectives divides the automation landscape into four zones: Automation ``Green Light'' Zone, Automation ``Red Light'' Zone, R\&D Opportunity Zone, and Low Priority Zone.} \textbf{a}, Tasks from \name are plotted in this desire-capability landscape. %
    \textbf{b}, We collect Y Combinator (YC) companies and map them to tasks based on the description on their official YC detail pages using \texttt{gpt-4.1-mini}. The average number of YC companies per task shows no significant difference across zones, highlighting the importance of steering more investment toward the Automation ``Green Light'' Zone and R\&D Opportunity Zone. \textbf{c}, We collect AI agent research papers from arXiv and evaluate their applicability to each occupational task in our database using \texttt{gpt-4.1-mini}. Encouragingly, the paper-task mappings are concentrated more in the R\&D Opportunity Zone, though increased emphasis on this area remains desirable.
    }
    \label{Fig:automation_readiness}
\end{figure*}

\subsection{Desire-Capability Landscape for AI Agents in the Workplace}
\label{sec:desire_capability}

\paragraph{Contrasting worker and AI expert perspectives delineate four task zones.}
While workers' preferences offer valuable guidance for socially beneficial AI agent deployment, delivering impact ultimately depends on aligning those preferences with technical feasibility. To investigate this, we jointly consider the worker-rated automation desire $A_w(t)$ and expert-assessed technological capability $A_e(t)$, visualized as a desire-capability landscape in \reffig{Fig:automation_readiness} \textbf{a}. 
This landscape divides into four zones:

\begin{enumerate}\setlength\itemsep{0em}
    \item {\textbf{Automation ``Green Light'' Zone}: Tasks with both high automation desire and high capability. These are prime candidates for AI agent deployment with the potential for broad productivity and societal gains.}
    \item {\textbf{Automation ``Red Light'' Zone}: Tasks with high capability but low desire. Deployment here warrants caution, as it may face worker resistance or pose broader negative societal implications.}
    \item {\textbf{R\&D Opportunity Zone}: Tasks with high desire but currently low capability. These represent promising directions for AI research and development.}
    \item {\textbf{Low Priority Zone}: Tasks with both low desire and low capability. These are less urgent for AI agent development.}
\end{enumerate}

Tasks from \name are broadly distributed across the landscape, with no strong correlation between $A_w(t)$ and $A_e(t)$ (Spearman $\rho=0.17$, $p<1e-6$). Overall, automation desire shows a negative correlation with both job loss concern (Spearman $\rho=-0.22$, $p<5e-11$) and enjoyment (Spearman $\rho=-0.28$, $p<3e-17$). This suggests both alignments and misalignments between worker desires and technological capabilities, with a consistent pattern: workers are less inclined to have AI agents automate tasks they enjoy or feel vulnerable about potential job loss, consistent with findings from prior literature~\citep{armstrong2024automationworkersperspective,GODOLLEI2023100342}.

\paragraph{Mapping investment to the desire-capability landscape reveals critical mismatches.}
To better understand where current investments are concentrated, we used Y Combinator (YC) companies\footnote{\url{https://www.ycombinator.com/companies}} as a proxy and mapped them to the tasks in our database using \texttt{gpt-4.1-mini} (one company could be mapped to multiple tasks, see Appendix~\ref{appendix:yc_company} for details). As shown in \reffig{Fig:automation_readiness} \textbf{b}, the company-task mappings are relatively evenly spread across the four zones. Most mapped tasks are concentrated in occupations related to software development and business analysis, with the top five occupations being: Computer and Information Systems Managers, Computer Programmers, Computer Systems Engineers/Architects, Software Quality Assurance Analysts and Testers, and Business Intelligence Analysts. 41.0\% of YC companies are mapped to Low Priority and Automation ``Red Light'' Zone; while many promising tasks within the ``Green Light'' Zone and Opportunity Zone remain under-addressed by current investments.

\paragraph{AI agent research papers  show an emphasis on the R\&D Opportunity Zone, but remain concentrated on a limited set of tasks.}
Following a similar methodology, we gathered research papers related to AI agents from arXiv\footnote{\url{http://arxiv.org}} and analyzed their alignment with various tasks to determine the distribution of research efforts.
\reffig{Fig:automation_readiness} \textbf{c} shows that research papers are more concentrated in the R\&D Opportunity Zone. While encouraging, the focus remains largely confined to computer science and engineering domains. The top three tasks covered are: (1) Computer and Information Research Scientists: Apply theoretical expertise and innovation to create or apply new technology, such as adapting principles for applying computers to new uses (1,169 papers); (2) Computer and Information Research Scientists: Analyze problems to develop solutions involving computer hardware and software (1,132 papers); (3) Computer Programmers: Perform or direct revision, repair, or expansion of existing programs to increase operating efficiency or adapt to new requirements (1,109 papers). These findings highlight the need to expand research efforts beyond a few domains to better support tasks in the R\&D Opportunity Zone, ensuring that future AI agents address a wider range of high-desire, high-impact opportunities.

\begin{figure*}[t]
    \centering
    \resizebox{\textwidth}{!}{%
    \includegraphics{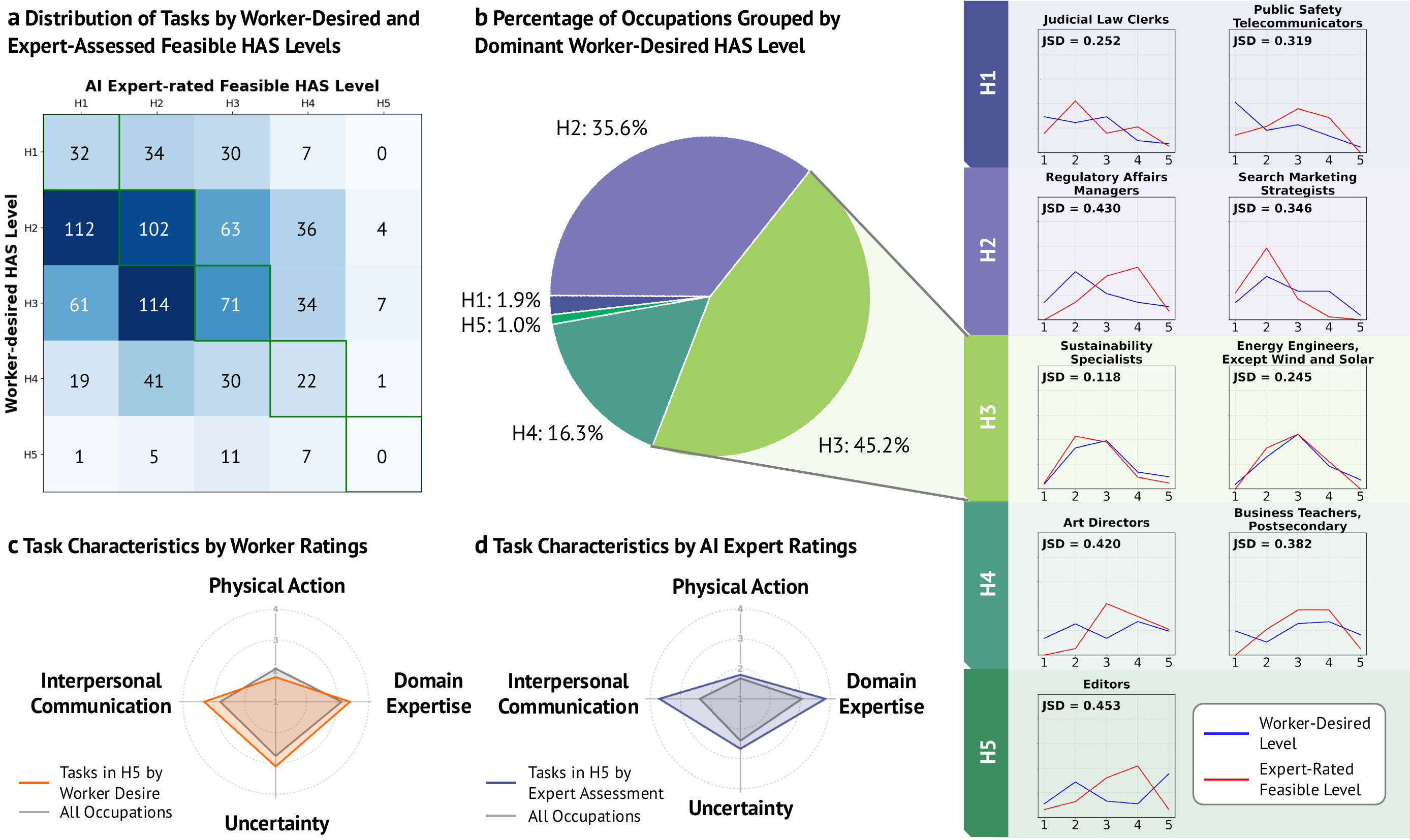}
    }
    \caption{\textbf{Distributions on the Human Agency Scale (HAS) reveal diverse patterns of AI agent integration across occupations and underscore opportunities for human–agent collaboration.} \textbf{a}, Comparison between worker-desired HAS levels ($H_w(t)$) and expert-assessed feasible HAS levels ($H_e(t)$) shows that workers generally prefer higher levels of human agency than what experts deem technologically necessary. \textbf{b}, Distribution of occupations by their dominant worker-desired HAS level shows that most occupations cluster around H3 and helps identify occupations that are at the poles of the agency spectrum. Each subplot displays the task-level distributions of worker-desired and expert-assessed HAS levels for a given occupation. Jensen-Shannon divergence (JSD) quantifies the difference between these distributions (see top 10 in \reftab{table:jsd_comparison}). Full occupation-level results are shown in \reffig{Fig:has_full_results}. \textbf{c}, Radar plot of task characteristics based on worker ratings indicates that tasks in the H5 region are particularly associated with Interpersonal Communication. \textbf{d}, Radar plot based on expert ratings shows that tasks in H5 are marked by strong Interpersonal Communication and Domain Expertise components.
    }
    \label{Fig:has_overview}
\end{figure*}

\subsection{Human Agency Scale (HAS) Spectrum}
\label{sec:has_result}

Beyond automation, AI agents hold promise for augmenting human work. To understand where and how this augmentation may occur, we analyze the distribution of both worker-desired HAS levels ($H_w(t)$) and expert-assessed feasible HAS levels ($H_e(t)$) across tasks within each occupation.

\paragraph{Where do worker desires and expert assessments diverge most on the Human Agency Scale?}
Each task in \name is assigned a worker-desired and expert-rated HAS level via majority vote. Among 844 tasks, 26.9\% receive matching levels from workers and AI experts. \reffig{Fig:has_overview} \textbf{a} shows workers generally prefer higher levels of human agency than what experts deem technologically necessary, with 47.5\% of tasks fall into the lower triangle of the matrix. To quantify this divergence, we compute the Jensen-Shannon Distance (JSD) between the distributions of $H_w(t)$ and $H_e(t)$T at the occupation level. Disagreements are most pronounced in the lower HAS range as \reftab{table:jsd_comparison} shows that five of the ten occupations with the highest JSD scores are also those that experts rate as H1 dominant. This signals potential frictions as AI adoption progresses. Ensuring socially responsible deployment of AI agents and supporting workers currently performing low-HAS tasks warrant further scrutiny.

\paragraph{What are the common patterns across the Human Agency Scale spectrum?}
As illustrated in \reffig{Fig:has_overview} \textbf{b} (with full results in \reffig{Fig:has_full_results}), many occupations (\eg, ``Sustainability Specialists'', ``Energy Engineers'') exhibit an inverted-U shaped distribution for both $H_w(t)$ and $H_e(t)$. While this trend in expert assessments might reflect current technological limitations---\ie, AI agents are not yet capable of fully replacing human involvement in most tasks---it is notable that workers in many domains also prefer a balanced, collaborative partnership with AI. H3 emerges as the dominant worker-desired level in 47 out of 104 occupations analyzed.

\paragraph{Which occupations stand out on the Human Agency Scale?}
Beyond the general inverted-U trend, we examine occupations at the extremes of the HAS spectrum. Lower HAS levels signify areas of greater potential AI exposure. According to AI experts' ratings, 16 out of 104 occupations are predominantly H1, even based on current capability estimates. These include roles such as ``Computer Programmers'', ``Proofreaders and Copy Makers'', and ``Travel Agents''. Occupations within the same sector also exhibit distinct trends in HAS levels. For example, ``Computer Programmers'' and ``Information Technology Project Managers'' display markedly different distributions (H1 vs. H4) when assessed by AI experts. Compared to \citet{eloundou2023gptsgptsearlylook}, which provides an early analysis of LLMs’ labor market impact and finds higher-wage occupations to be more exposed, our results show that while most occupations in \name are indeed exposed to AI agents and do not fall into H5 (essential human involvement), those involving more routine tasks and easily verifiable outcomes tend to require lower human agency.

Very few occupations are characterized by a dominant HAS Level 5 (indicating essential human involvement). ``Editors'' is the only occupation where workers predominantly desire H5. According to AI expert assessments, only ``Mathematicians'' and ``Aerospace Engineers'' fall into this category. Representative work descriptions from worker transcripts for these occupations are provided in Appendix~\ref{appendix:audio_response}. We further investigate what distinguishes H5 tasks. Among the four quantitative dimensions in our auditing framework (\reffig{Fig:audit_framework}), worker ratings highlight Interpersonal Communication as a defining feature of H5 tasks (\reffig{Fig:has_overview} \textbf{c}), while expert ratings emphasize both Interpersonal Communication and Domain Expertise (\reffig{Fig:has_overview} \textbf{d}).

\paragraph{What forms of human-agent collaboration do workers envision?}
\reffig{Fig:has_overview} \textbf{b} suggests strong potential for collaborative AI. Our analysis of audio transcripts supports this: the vast majority of workers express either a desire or openness to collaborating with AI to enhance their work. Analyzing these narratives further illustrates how workers concretely envision human–agent partnerships. The most common paradigm is ``role-based'' AI support (described by 23.1\% of workers), where individuals anticipate \emph{utilizing AI systems that embody specific roles or personalized functions} (\emph{``[I would like to have an AI agent] trained to automatically analyze the quality control reports of raw sequencing data (e.g., FastQC output) and flag potential issues with specific samples or sequencing runs, [...] and suggesting appropriate pre-processing steps.'' ``It's just for me to set up my AI.''}). Furthermore, 23.0\% of workers express a desire for AI systems to function as a supportive assistant for some or all aspects of their workflow (\emph{``[I envision the AI agent] as an assistant who is doing research for me. However, I review every answer because we cannot rely on its accuracy.''}), while 16.5\% mention pure automation by AI for some aspects of their workflow.

\begin{figure*}[t]
    \centering
    \resizebox{\textwidth}{!}{%
    \includegraphics{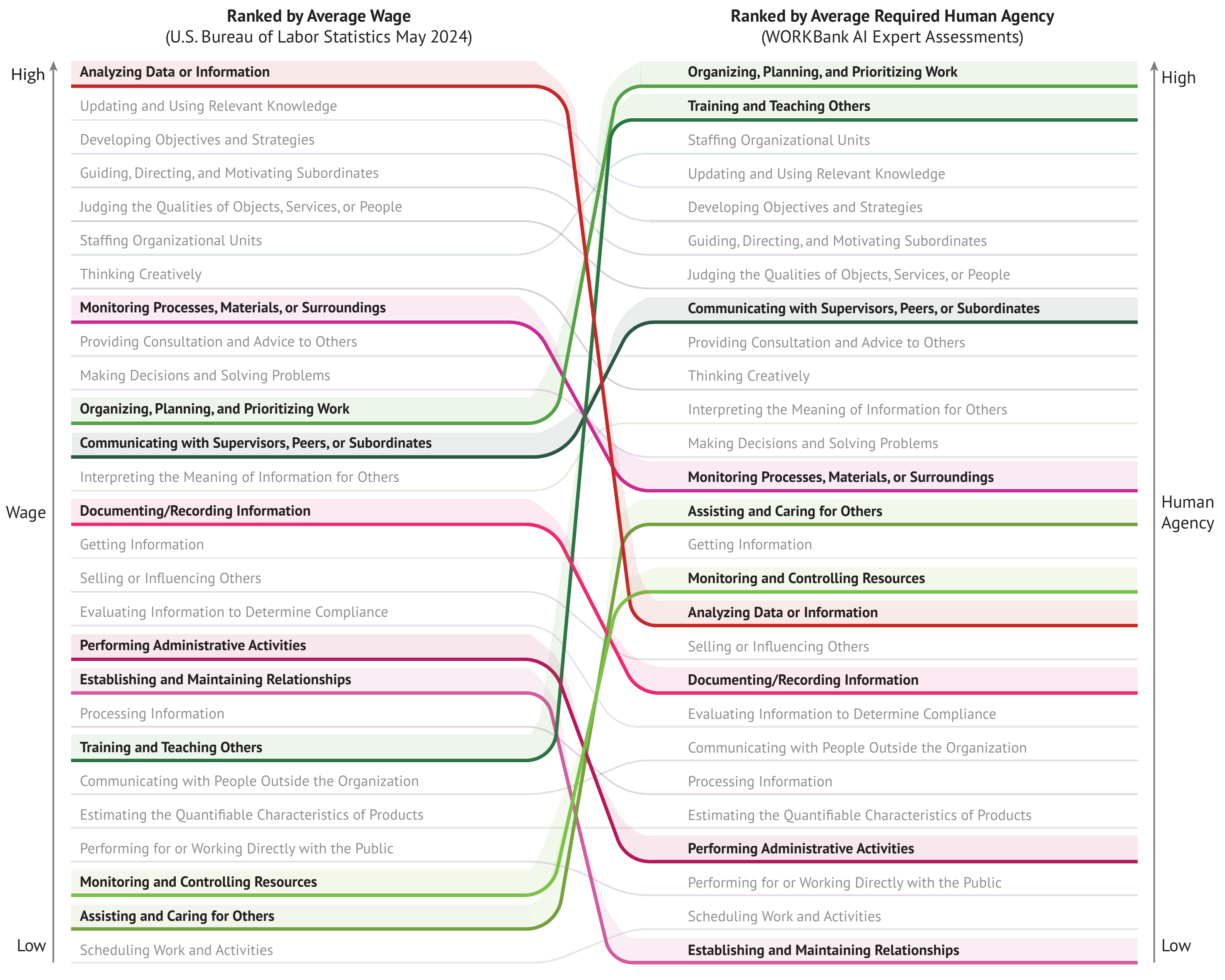}
    }
    \caption{
    \textbf{Comparing skill rankings by average wage and required human agency.} Each line represents a skill (Generalized Work Activity) mapped from O*NET tasks. Based on the skill-task mapping, we compute the average wage using data from the U.S. Bureau of Labor Statistics (May 2024) and the average expert-assessed human agency level $H_e(t)$ to indicate the degree of human involvement required as AI agents enter the workplace. Skills are ranked by average wage (left) and average required human agency (right). The figure highlights the top five skills with the largest upward (green) and downward (red) shifts in rank, suggesting a potential shift in valued workplace skills---from information processing toward interpersonal and organizational competencies. See Appendix~\ref{appendix:skill} for skill analysis details and full skill descriptions.
    }
    \label{Fig:skill_shift}
\end{figure*}

\subsection{The Potential Shift of Core Human Skills}

Variation in Human Agency Scale (HAS) across occupations suggests that certain types of human work are more susceptible to automation, while others hold greater potential for augmentation. To this end, we examine the characteristics of tasks that require high human agency to understand how AI agents may shift skill demands. 

Translating task-level changes into implications for education and skill training has long been a key lens for analyzing technological transformation, notably pioneered by \citet{autor2003skill} in the wave of computers. To operationalize this lens, we align each task with its related skills (Generalized Work Activities) as defined by O*NET (Appendix~\ref{appendix:skill}). For example, the task ``Financial Managers: Approve, reject, or coordinate the approval or rejection of lines of credit or commercial, real estate, or personal loans'' will be mapped to ``making decisions and solving problems'' and ``guiding, directing, and motivating subordinates''. We compute the average expert-assessed human agency level $H_e(t)$ for each skill to estimate the degree of human involvement required as AI agents enter the workplace. We also compute the average wage for each skill, using data from the U.S. Bureau of Labor Statistics (Appendix~\ref{appendix:wage}). Wage serves as a proxy for the current economic value of each skill~\citep{dey2019job}.

As shown in \reffig{Fig:skill_shift}, by comparing skill rankings by average wage and required human agency, our analysis reveals three emerging trends:

\begin{enumerate}
    \item {\textbf{Shrinking demand for information-processing skills.} Skills related to analyzing data and updating knowledge---while common in today’s high-wage occupations (as shown in the left side of \reffig{Fig:skill_shift} in red color)---are less prominent in tasks that demand high human agency.}
    \item {\textbf{Greater emphasis on interpersonal and organizational skills.} Skills involving human interaction, coordination, and resource monitoring are more frequently associated with high-HAS tasks (as shown in the left side of \reffig{Fig:skill_shift} in green color), even if they are not currently prioritized in wage-based evaluations.}
    \item {\textbf{High-agency skills span diverse aspects.} The top 10 skills with the highest average required human agency encompass a broad range, from interpersonal and organizational abilities to decision-making and quality judgment.}
\end{enumerate}

These findings provide early signals of how AI agent integration may reshape core occupational competencies. As workplace AI agents continue to evolve, longitudinal tracking of task-level changes could yield further insights into how human roles and required skills evolve.

%% file: 2-Related_Work.tex
\noindent
\textbf{Digital AI Agents}\quad
The aspiration to build AI agents capable of dynamically directing their own processes to accomplish complex goals dates back to the early days of artificial intelligence~\citep{mccarthy1959programs,Genesereth+Nilsson:1987}. Recent advances in foundational models, particularly large language models (LLMs), have sparked a surge in the development of digital AI agents that leverage LLMs to plan actions and interface with external tools~\citep{sumers2023cognitive,wang2024survey}. These agents have demonstrated the ability to carry out workflows across diverse domains, including software engineering~\citep{yang2024swe,openhands}, analytical writing~\citep{shao-etal-2024-assisting,jiang-etal-2024-unknown}, and customer support~\citep{yao2024taubench}.

While many of these agents are designed for full automation, they can also be structured to collaborate with humans. Collaborative Gym~\citep{shao2024collaborative} pioneered the concept of human-agent collaboration, demonstrating that for certain tasks, joint human-agent performance can surpass that of fully autonomous agents, even when those agents are capable of completing the tasks independently. This underscores the potential of AI agents to augment, rather than simply replace, human labor~\citep{brynjolfsson2022turing}. The auditing framework proposed in this work systematically examines augmentation versus automation by introducing the Human Agency Scale (HAS), which evaluates the level of ideal human involvement across different workflows.

One limitation of prior work on AI agents is its frequent focus on a narrow set of domains. Existing benchmarks, such as GAIA~\citep{mialon2023gaia}, AgentBench~\citep{liu2023agentbench}, OSWorld~\citep{xie2024osworld}, while valuable for assessing agent capabilities, often rely on task collections that are curated in a constrained manner. Such approach, while useful for capability evaluation, fails to provide a holistic and worker-centric understanding of how these agents could be integrated into the broader workforce. By sourcing tasks from the U.S. Department of Labor's O*NET database, our work provides a more comprehensive and systematic understanding of the potential landscape for digital AI agents.

\noindent
\textbf{The Economic Impacts of Generative AI}\quad
A broad body of work in digital economics has examined the implications of AI, spanning from early machine learning models~\citep{brynjolfsson2017can} and computer vision systems~\citep{svanberg2024beyond} to the recent surge of large language models (LLMs) and generative AI~\citep{demirci2025ai,eloundou2023gptsgptsearlylook,handa2025economic,hoffmann2024generative}. Following the launch of ChatGPT, \citet{eloundou2023gptsgptsearlylook} provided an early analysis of LLMs’ potential labor market impact, estimating that approximately 80\% of the U.S. workforce has at least some tasks exposed to LLM capabilities. However, their analysis did not incorporate the dimension of worker desire and focused primarily on LLM via ChatGPT or the OpenAI playground rather than the broader scope of AI agents. More recent work leveraging real user data from Claude.ai, a state-of-the-art LLM chatbot, to identify which economic tasks users actually perform with AI~\citep{handa2025economic}. In parallel, field studies in customer support organizations have shown that AI-assisted chatbots can improve worker productivity~\citep{brynjolfsson2025generative}. As AI agents continue to evolve beyond standalone LLM chatbots, our study provides an early audit of their readiness for workplace integration.%

%% file: 8-Discussion.tex
Advancements in AI agents are unlocking a wide range of possibilities that may fundamentally reshape the workplace. This paper presents the first large-scale audit of both worker desire and technological capability for AI agents in the context of automation and augmentation. Based on data collected between January and May 2025, we construct the \name database and find that domain workers generally express positive attitudes toward AI agent automation, particularly for repetitive and low-value tasks.

By integrating both worker and expert perspectives, we introduce the automation desire–capability landscape, which offers actionable insights for prioritizing AI agent research and investment. Besides the traditional automate-or-not dichotomy, our Human Agency Scale (HAS) uncovers diverse patterns of AI integration across occupations, with a dominant inverted-U trend that underscores the potential for human–agent collaboration.

Beyond informing AI agent research and deployment strategies, our findings also have implications for workforce development. As AI agents reshape the demand for core human skills, our findings suggest that examining strategies for worker reskilling and retraining is a valuable direction for future research.

\paragraph{Limitations}
While our audit offers a comprehensive snapshot of worker perspectives and technological capabilities of workplace AI agents, several limitations should be considered:

First, our quantitative assessments are grounded in existing occupational tasks defined by the O*NET database, which does not account for new tasks that may emerge as AI agents become more integrated into the workplace. Further analysis of the open-ended worker transcripts could uncover emerging task patterns and enrich our understanding of evolving occupational tasks.%

Second, although we guided participants to reflect on potential job loss and task enjoyment, domain workers may still lack full awareness of the evolving capabilities and limitations of AI agents~\citep{hazra2025ai}, potentially shaping their responses. We partially mitigate this limitation by including only occupations with at least 10 worker responses in the AI Agent \name. Robustness checks and further discussion are provided in Appendix~\ref{appendix:robustness}, and complement it with AI experts' assessment.  

From an incentive perspective, some workers may also withhold honest feedback due to concerns about job security or surveillance. 
We recognize that this is a real concern, which is why we're committed to a worker-focused approach that surfaces real concerns and co-designs systems that reflect workers’ values. We see such resistance as a critical input that helps guide responsible deployment. By prioritizing workers' perspectives, we enable workers to %
play an active role in shaping the future of work, rather than just adapting to it.

Third, the current version of AI Agent \name only covers 104 occupations, a subset of the 287 computer-using occupations identified with the O*NET database. In our study, we launched the survey interface on Prolific, Upwork, and LinkedIn in January 2025 and concluded data collection in May 2025 to ensure temporal consistency. These 104 occupations were retained after filtering for adequate representation ($\geq 10$ participants per occupation). 
While our database exhibits strong coverage and demographic representativeness (see Appendix~\ref{appendix:statistics}), our findings may not cover the full picture of AI agents for the workplace.

Finally, the AI Agent \name reflects the present state of generative AI and agentic systems as of early 2025. As AI capabilities continue to evolve, the landscape of feasible and desirable agent-supported tasks will likely shift. While our framework offers a timely and structured baseline, future iterations of this audit will be essential for tracking long-term trends and informing the responsible development of workplace AI systems.

%% file: 9999-Acknowledgement.tex
We would like to thank Chuchu Jin and Yanzhe Zhang for their help in distributing the survey interface on Linkedin. We are grateful to Hao Zhu for database setup, to Will Held, Dora Zhao, Omar Shaikh, Yifan Mai, Sunny Yu, Zachary Robertson for their valuable feedback on the manuscript, and to all members of Stanford SALT lab for their suggestions at different stages of this project. This work would not have been possible without the thoughtful participation of the 1,500 domain workers and 52 AI experts who contributed to the study. This research is supported in part by grants from ONR grant N000142412532, and NSF grant IIS-2247357 and support from the Stanford Digital Economy Lab.

%% file: 999-Appendix.tex
\section{Survey Details}
\label{appendix:survey_details}

We instantiate our auditing framework (see \refsec{sec:framework}) with an audio-enhanced, semi-structured survey interface to collect first-hand data from domain workers who actually perform those tasks. Our survey is structured as follows:

\begin{itemize}
    \item {
    A \textbf{Mini-interview Section} designed to explore participants’ work process and perspectives on the role of AI agents in their work. This section consists of five open-ended questions, allowing participants to share their thoughts freely and edit their audio transcripts in real time:
    \begin{itemize}
        \item [A1] Could you please briefly describe what you do for your work?
        \item [A2] What tasks do you typically do for your work? Think about this question based on the time you spend on each of them.
        \item [A3] Please tell us more about the tools or software you use for these tasks. Please try to sort them by usage frequency.
        \item [A4] For the three tasks that you spend the most of your time on, could you walk us through your process of completing each of them? 
        \item [A5] How do you envision using AI in your daily work?
    \end{itemize}
    }
    \item {
    A \textbf{Task Rating Section} assessing both automation desire ($A_w(t)$) and the desired Human Agency Scale level ($H_w(t)$) for tasks associated with the participant's occupation. For each task, participants respond to a series of structured questions. Items T.I3, T.A1–A3, and T.C1–C5 are all rated on a 5-point Likert scale:

    \quad\quad\textsc{Task Familarity questions:}
    \begin{itemize}
        \item [T.I1] Have you done this task before? \\
        \textit{(Yes/No; if ``No'', remaining questions are skipped.)}
    
        \item [T.I2a] With respect to this task, I consider myself a... \\
        \textit{(Novice, Average, Expert)}
    
        \item [T.I2b] How much time do you spend on this task in your daily work schedule? \\
        \textit{(10\%, 30\%, 50\%, 70\%, 100\%)}
    
        \item [T.I3] How closely is this task related to your core skills or unique strengths that are essential to your job? 
    
     \textsc{Automation Desire rating questions:}
        \item [T.A1] If an AI can do this task for you completely, how worried would you be that your job will be replaced? 
    
        \item [T.A2] Without thinking about salary, how much do you enjoy doing this task? 
    
        \item [T.A3] If an AI can do this task for you completely, how much do you want an AI to do it for you?
    
        \item [T.A4] Why would you like this task to be automated by AI? (Shown if T.A3 $\geq$ 3; multi-select options):
        \begin{itemize}
            \item Automating this task would free up my time for higher-value work.
            \item This task is repetitive or tedious.
            \item Automating this task would improve the quality of my work.
            \item The task is stressful or mentally draining.
            \item This task is complicated or difficult.
            \item Automating this task would help me scale and handle higher output.
        \end{itemize}

    \textsc{Human Agency Scale rating questions}:
    
        \item [T.C1] How much does this task require taking physical actions or physical labor? 
    
        \item [T.C2] How much does this task require dealing with uncertainty or making high-stake decisions?
    
        \item [T.C3] How much does this task require specific domain expertise (such as specialized knowledge, unspoken wisdom, or insights gained through experience)?
    
        \item [T.C4] How much does this task depend on interpersonal communication or empathy?
    
        \item [T.C5] If AI were to assist in this task, how much of your collaboration would be needed to complete this task effectively? (References to H1 to H5 are provided)
    
        \item [T.C6] Why would collaboration be needed for this task? (Shown if T.C5 $\geq$ 3; multi-select options):
        \begin{itemize}
            \item This task requires physical actions.
            \item This task involves making high-stake decisions which I would like to control.
            \item This task requires specific domain knowledge.
            \item The task involves nuanced communication or interpersonal skills.
            \item The task needs validation or oversight to ensure quality.
            \item The task is dynamic and requires adapting to changing circumstances.
            \item The task has ethical, sensitive, or subjective aspects.
    
        \end{itemize}
    \end{itemize}
    }
    \item{A \textbf{Demographic Question Section} where we ask about information on age, gender, race, income, education, years of experience in occupation, attitude towards AI, zip code, political orientation, and details about their familiarity with LLMs and how they currently use them (\ie, types of usage and frequency).}
\end{itemize}

\section{Robustness Analysis}
\label{appendix:robustness}

\subsection{Annotation Agreement of AI Expert Assessments}

AI experts with practical R\&D experience provided ratings for current automation capability $A_e(t)$ and feasible human‐agency levels $H_e(t)$. To ensure high‐quality annotations, we applied the following controls:
\begin{itemize}
  \item \textbf{Expert qualifications.} Each expert satisfied at least one of:
    \begin{enumerate}
      \item Current PhD student specializing in NLP, large language models, or AI agents.
      \item PhD in Computer Science with demonstrated expertise in AI, LLMs, or agentic systems.
      \item Industry practitioner (\eg, machine learning engineer or research scientist) with hands‐on experience in LLMs and agentic systems.
    \end{enumerate}
    In total, 52 experts were recruited from institutions including Stanford University, MIT, Google, and xAI, \etc.
  \item \textbf{Assessment protocol.} Every task $t$ was independently assessed by at least two experts, with additional reviews to ensure that $A_e(t)$ and $H_e(t)$ ratings exhibit a standard deviation $\leq1$.
\end{itemize}

Inter‐annotator agreement, measured by Krippendorff’s $\alpha$, was 0.539 for $A_e(t)$ and 0.511 for $H_e(t)$.

\subsection{Mixed-Effects Model Regression on Worker Responses}

\begin{table*}
\centering
\resizebox{\textwidth}{!}{%
\input{tables/mixedlm_summary}
}
\caption{Fixed‐effects estimates from the mixed‐effects regression predicting automation desire ratings.}
\label{table:mixedlm_summary}
\end{table*}

To assess the extent to which workers’ automation desire ratings reflect intrinsic task properties rather than individual demographics, we fitted a linear mixed‐effects model of the form
\begin{align*}
y_{ij} &= \beta_0 + \sum_{k=1}^{K} \beta_k\,X_{k,ij} + u_{j} + \varepsilon_{ij},\\
u_{j} &\sim \mathcal{N}(0,\sigma_u^2), \quad
\varepsilon_{ij} \sim \mathcal{N}(0,\sigma_\varepsilon^2),
\end{align*}
where \(y_{ij}\) is the automation desire rating provided by worker \(i\) on task \(j\), \(X_{k,ij}\) are the \(K\) demographic and attitude covariates (age, gender, education, experience, LLM familiarity/use, income, political affiliation, LLM-usage subtypes, and AI-attitude scales), \(\beta_k\) their fixed effects, \(u_j\) a task‐specific random intercept, and \(\varepsilon_{ij}\) the residual error.  The model was estimated by REML using the \texttt{statsmodels} MixedLM interface in Python.

The fitted model (see \reftab{table:mixedlm_summary}) yielded the following variance‐component estimates:
\[
\widehat{\sigma}_u^2 = 0.066,\qquad
\widehat{\sigma}_\varepsilon^2 = 1.254,
\]
implying an intraclass correlation
\[
\mathrm{ICC} \;=\;\frac{\widehat{\sigma}_u^2}{\widehat{\sigma}_u^2+\widehat{\sigma}_\varepsilon^2}
\;=\;0.050,
\]
\ie, roughly 5\% of total variance in automation desire is attributable to between‐task differences.  This ``small‐to‐moderate'' ICC confirms that task‐level properties carry a real signal in workers’ automation desire ratings.

Among fixed effects, higher educational attainment (``Doctorate'' vs. ``Bachelor’s'': \(\hat\beta=0.236\), \(p=0.037\)) and greater work experience (``>10 years'' vs. ``1–2 years'': \(\hat\beta=0.229\), \(p<0.01\)) were associated with increased automation desire.  Attitudinal scales also showed significant associations: Strong agreement that “AI relieves tedious work” predicted higher desire (\(\hat\beta=0.685\), \(p<0.001\)), while stronger “AI suffering” attitudes predicted lower desire (\(\hat\beta=-0.438\), \(p<0.001\)).  %
Income levels were positively related to automation desire (\eg ``\$529K+'' vs. ``\$0–30K'': \(\hat\beta=0.687\), \(p<0.001\)).%

We further conduct an additional analysis focusing on tasks where the task-level signal is stronger. We identified ``high-signal'' tasks based on two criteria: (1) low within-task variance in worker ratings (below the median variance), and (2) strong separate from the overall mean desire rating (above the median absolute deviation). This filtering retained 238 tasks. Fitting on this high-signal subset, the same linear mixed-effects model yields an ICC of 0.405. Critically, the distribution of tasks across the four zones remained remarkably stable: in the high-signal subset, 41.60\% of tasks fall in the Automation ``Green Light'' Zone (vs. 43.13\% in the full dataset), 31.09\% in the Automation ``Red Light'' Zone (vs. 31.64\%), 15.13\% in the Low Priority zone (vs. 12.56\%), and 12.18\% in the R\&D Opportunity zone (vs. 12.68\%). Together, these results indicate that, after controlling for a broad array of individual differences, task identity still explains a meaningful fraction of variance in automation desire. In the main paper, we use the average ratings for analysis.

\section{Usage of External Data and Resources}
\subsection{Occupational Information Network (O*NET)}
\label{appendix:task_source}

We source occupational tasks in this study from O*NET (version 29.2) Task Statements\footnote{https://www.onetcenter.org/dictionary/29.2/excel/task\_statements.html}. The O*NET database is a regularly updated database containing information about occupations across the United States. O*NET maps occupations to knowledge, skills, and abilities on different levels of granularity, as well as to tasks and detailed work activities. In O*NET, tasks are specific work activities that can be unique for each occupation. %
In total, there are 18,796 task statements spanning across 923 occupations.

Each task statement is associated with O*NET-SOC Code, Title (\ie, occupation), and Task Type (\ie, ``Core''/``Supplementary''). Moreover, O*NET provides annotations of task categories based on the frequency of the task in seven categories (``Yearly or less'', ``	More than yearly'', ``More than monthly'', ``More than weekly'', ``Daily'', ``	Several times daily'', ``	Hourly or more'')\footnote{https://www.onetcenter.org/dictionary/29.2/excel/task\_categories.html}. As this work focuses on digital AI agents, we filter these task statements based on the following criteria:

\begin{enumerate}
    \item {The occupation mainly involves using computers in its work as judged by \texttt{gpt-4o}.}
    \item {The task can be finished on the computer as judged by \texttt{gpt-4o}}
    \item {The task does not miss annotation for ``Core''/``Supplementary''.}
    \item {The task will be done more than monthly.}
\end{enumerate}

After filtering, there are 2,131 tasks remaining, spanning across 287 occupations.

\input{prompts/occupation_use_computer}

\input{prompts/workflow_use_computer}

\subsection{Occupational Employment and Wage Statistics by U.S. Bureau of Labor Statistics}
\label{appendix:wage}

We use occupational employment and wage statistics from the U.S. Bureau of Labor Statistics (BLS) to contextualize our findings with economic data. Specifically, we draw on data from the BLS May 2024 Occupational Employment and Wage Statistics Query System\footnote{\url{https://www.bls.gov/oes/tables.htm}} to obtain the ``Annual Mean Wage'' and ``Employment'' (\ie, number of employees) fields for each occupation in our database. These fields, combined with the collected first-hand data in \name, inform the analysis presented in Figure~\ref{Fig:skill_shift}.

\subsection{Claude.ai Usage Data}

To shed light on the relationship between current large language model (LLM) usage and future worker desires, we compare WORKBank automation desires with existing Claude.ai usage data from the Anthropic Economic Index~\cite{handa2025economic}. The Anthropic dataset reports Claude usage at the task level, following O*NET task definitions. These standardized definitions allow us to directly map tasks from the Anthropic dataset to corresponding tasks in the WORKBank database, enabling a structured comparison across both sources. Our data shows that the top 10
occupations with the highest average automation desire represent only 1.26\% of total usage (\reffig{Fig:automation_desire} \textbf{c}).

\subsection{Y Combinator (YC) Company and AI Agent Research Paper Data}
\label{appendix:yc_company}

We collect data on Y Combinator (YC) companies and AI agent research papers to assess how current investment and research efforts align with the desire–capability landscape revealed by the \name database (\reffig{Fig:automation_readiness}). To enable this analysis, we developed an LLM-assisted pipeline that systematically identifies and maps relevant YC startups and academic publications to specific occupational tasks in the \name database.

\paragraph{YC Company Collection Process}
The full list of YC companies was retrieved on April 28, 2025, from the official YC website\footnote{\url{https://www.ycombinator.com/companies}}. The initial dataset comprised 5,156 companies. Company descriptions were collected from their respective YC detail pages and filtered using an LLM-based process (\texttt{gpt-4.1-mini}). Each company description was assessed using a binary classification prompt (see \hyperlink{yc-classifier}{Prompt for YC Company Classifier}) to determine whether the company is AI-relevant. This process identified 1,723 AI-related companies.

\paragraph{AI Agent Research Paper Collection Process}
Academic papers were obtained from the arXiv official website\footnote{\url{http://arxiv.org}}, with a submission cut-off date of April 24, 2025. Papers were first screened by keyword: their title or abstract must contain ``language model'' (case-insensitive) and either ``agent'' or ``system.'' This yielded an initial set of 17,064 papers. This set was refined with \texttt{gpt-4.1-mini} using a checklist (see \hyperlink{paper-classifier}{Prompt for AI Agent Paper Classifier}) to verify that each paper: (i) describes tasks extending \emph{beyond single-turn raw text completion}, (ii) presents an implemented pipeline, (iii) conducts task-level evaluations, and (iv) involves a realistic task scenario. This process produced a final selection of 1,222 papers. For each paper passing this filter, we performed an additional round of task extraction (see \hyperlink{paper-task-extractor}{Prompt for Paper Task Extractor}); the extracted task statement serves as the paper's representative description in the subsequent mapping step.

\paragraph{Task Mapping Process}
For each YC company and AI agent paper identified through the above processes, we again employed \texttt{gpt-4.1-mini} to perform binary classification of their applicability to each occupational task in the \name database. For YC company-to-task mapping, see \hyperlink{company-to-workflow-classifier}{Prompt for Company-to-Task Classifier}; for agent paper-to-task mapping, see \hyperlink{paper-to-workflow-classifier}{Prompt for Paper-to-Task Classifier}.

\input{prompts/workflow_mapping}

\clearpage
\section{\name Statistics}
\label{appendix:statistics}

As detailed in Section~\ref{sec:dataset}, we retained only occupations with at least ten worker responses from January to May 2025, yielding 1,500 individual assessments across 104 occupations. In this section, we present detailed statistics on occupational coverage, participant demographics, and recruitment sources in the \name database.

\subsection{Full List of Included Occupations}

\begin{table*}[h]
\centering
\resizebox{\textwidth}{!}{%
\input{tables/occupation-list}
}
\caption{List of occupations covered in the \name database and the number of survey participants from each occupation.}
\label{table:occupation_list}
\end{table*}

\clearpage

\subsection{Coverage of the U.S. Workforce}
\label{appendix:workforce_coverage}
We evaluate the representativeness of \name by comparing its sector-level distribution with U.S. workforce data from the Bureau of Labor Statistics\footnote{\url{https://www.bls.gov/oes/tables.htm}}. \reffig{Fig:sector_dist} presents a breakdown of workforce coverage by sector, contrasting our database with the full U.S. workforce (\textbf{a}) and with the workforce restricted to the 104 occupations included in our database (\textbf{b}). Overall, the comparisons suggest that our database captures a broad and representative cross-section of the U.S. workforce at the sector level.

\subsection{Domain Worker Demographic Information}

\begin{figure*}[t]
    \centering
    \resizebox{\textwidth}{!}{%
    \includegraphics{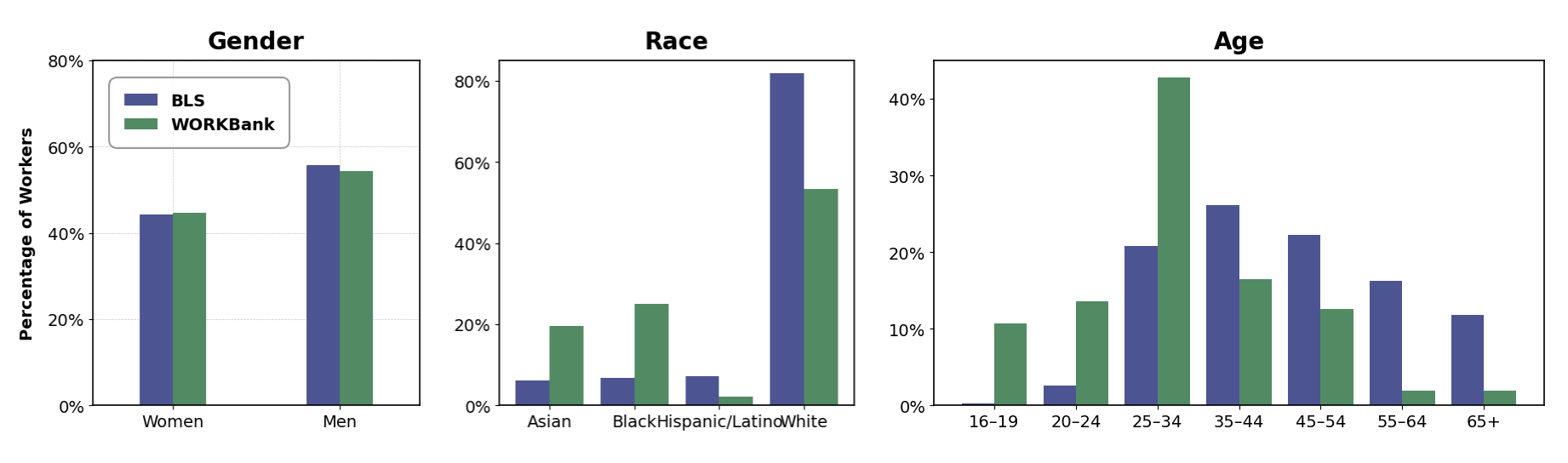}
    }
    \caption{Demographic distribution of worker participants in our study compared to U.S. workforce demographics for the same set of occupations covered in the \name database, based on data from the Bureau of Labor Statistics.}
    \label{Fig:demographics}
\end{figure*}

We compare the demographic profile of domain workers in \name with that of the U.S. workforce. U.S. workforce demographic data are sourced from the 2024 Annual Averages of the Bureau of Labor Statistics (BLS) Current Population Survey\footnote{\url{https://www.bls.gov/cps/tables.htm}}. To ensure a fair comparison, we filter the BLS data to include only the 104 occupations represented in our database. \reffig{Fig:demographics} shows the breakdown across key demographic dimensions. Our participant pool has a comprehensive demographic coverage, with a tendency toward younger age groups.

\subsection{Recruitment Sources}
To obtain worker assessments, we mainly recruited domain workers through Prolific, Upwork, and LinkedIn. \reftab{table:worker_distribution} reports the breakdown of recruited domain workers before data filtering. Fo expert assessments, we recruited 52 AI experts through word of mouth. These AI experts include PhD students, postdocs, professors, and industry practitioners spanning across different career stages. \reftab{table:expert_distribution} reports their career stage and institutional distribution.

\begin{table*}[h]
\centering
\resizebox{\textwidth}{!}{%
\input{tables/worker_distribution}
}
\caption{Breakdown of recruitment sources for domain workers in this study. The numbers shown reflect all participants recruited before data filtering. Breakdowns by recruitment source after data filtering are not available, as we did not link participant personal information with data points in WORKBank to preserve privacy.}
\label{table:worker_distribution}
\end{table*}

\begin{table*}[h]
\centering
\resizebox{\textwidth}{!}{%
\input{tables/expert_distribution}
}
\caption{Career stage and institutional distribution of 52 recruited AI experts.}
\label{table:expert_distribution}
\end{table*}

\clearpage
\section{Additional Results}
\subsection{Top 20 Tasks Workers Want Automated}

\begin{table*}[h]
\centering

\resizebox{\textwidth}{!}{%
\input{tables/top-20-automation-desire}
}
\caption{Top 20 tasks with highest average automation desire.}
\label{table:top_automation_desire}
\end{table*}

\clearpage

\clearpage
\subsection{Bottom 20 Tasks Workers Want Automated}

\begin{table*}[h]
\centering

\resizebox{\textwidth}{!}{%
\input{tables/bottom-20-automation-desire}
}
\caption{Bottom 20 tasks with lowest average automation desire.}
\label{table:bottom_automation_desire}
\end{table*}

\clearpage

\begin{figure*}[th]
    \centering
    \resizebox{0.9\textwidth}{!}{%
    \includegraphics{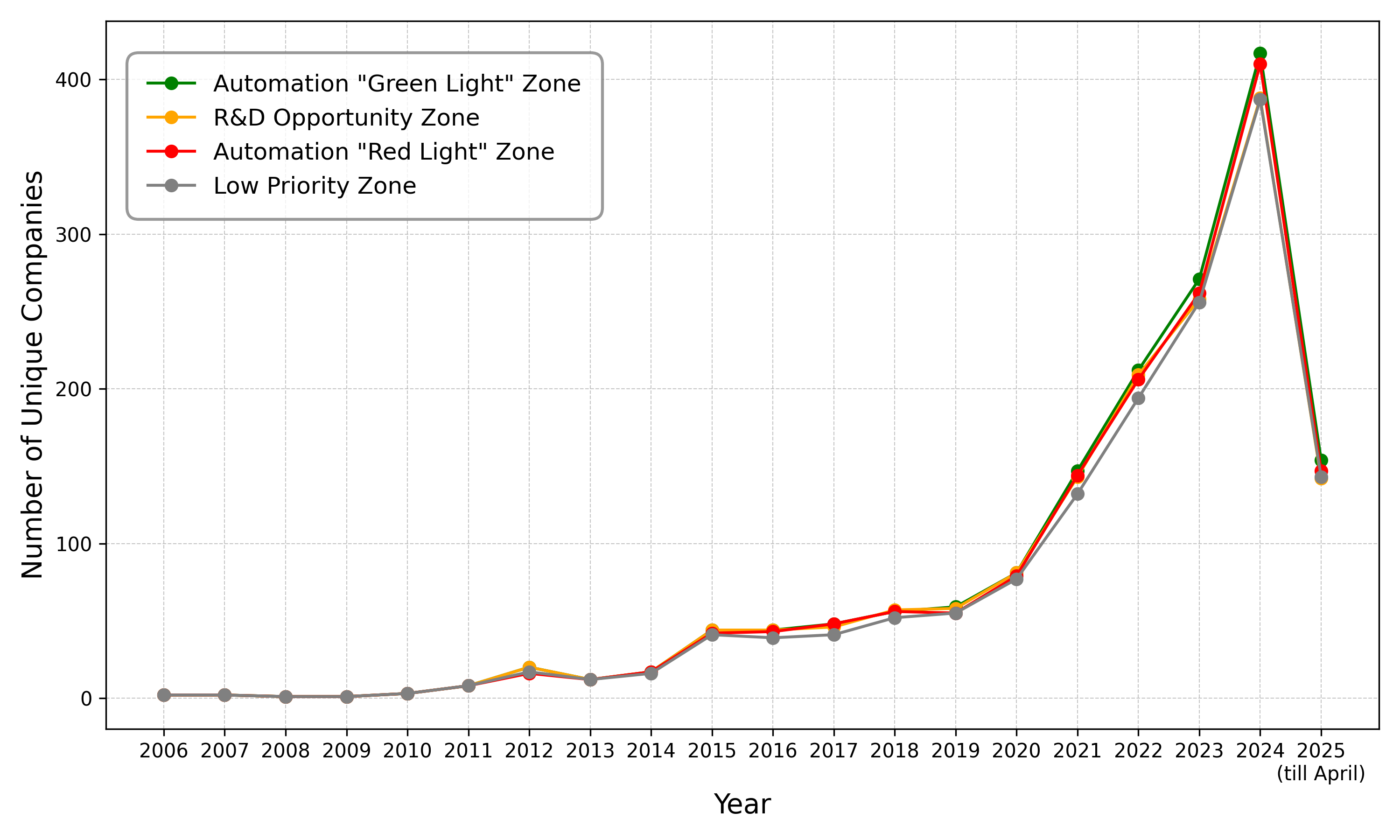}
    }
    \caption{\textbf{Number of unique newly funded Y Combinator companies mapped to each task zone in the automation desire–capability landscape (2006–2025, through April).} Despite the exponential growth in AI agent startups, the temporal trends across all four zones remain largely parallel. Notably, there is no disproportionate concentration in the Automation ``Green Light'' or R\&D Opportunity Zones---areas that warrant greater attention.%
    }
    \label{Fig:yc_company_trend}
\end{figure*}

\subsection{Y Combinator Investment Patterns Across Task Zones}

As discussed in~\refsec{sec:desire_capability}, jointly considering the worker-rated automation desire $A_w(t)$ and expert-assessed technological capability $A_e(t)$ divide tasks in \name into four zones: Automation ``Green Light'' Zone, Automation ``Red Light'' Zone, R\&D Opportunity Zone, Low Priority Zone

\reffig{Fig:yc_company_trend} illustrates the number of unique newly funded YC companies mapped to each task zone from 2006 to 2025 (till April). Despite the exponential growth in AI agent startups, the distribution across zones has remained relatively uniform over time. Notably, there is no disproportionate concentration in the Automation ``Green Light'' or R\&D Opportunity Zones---areas that warrant greater attention. While the task zone classifications reflect a static snapshot and may not reflect the status in the past, the findings nonetheless suggest a misalignment between where investments are flowing and the joint perspective of both those developing the technology and the workers the technology shall aim to support.

\clearpage
\thispagestyle{empty}
\newgeometry{bottom=0cm}
\subsection{Full Human Agency Scale Results}
\begin{figure*}[h]
  \centering
  \begin{minipage}{\textwidth}
    \centering
    \includegraphics[width=\textwidth]{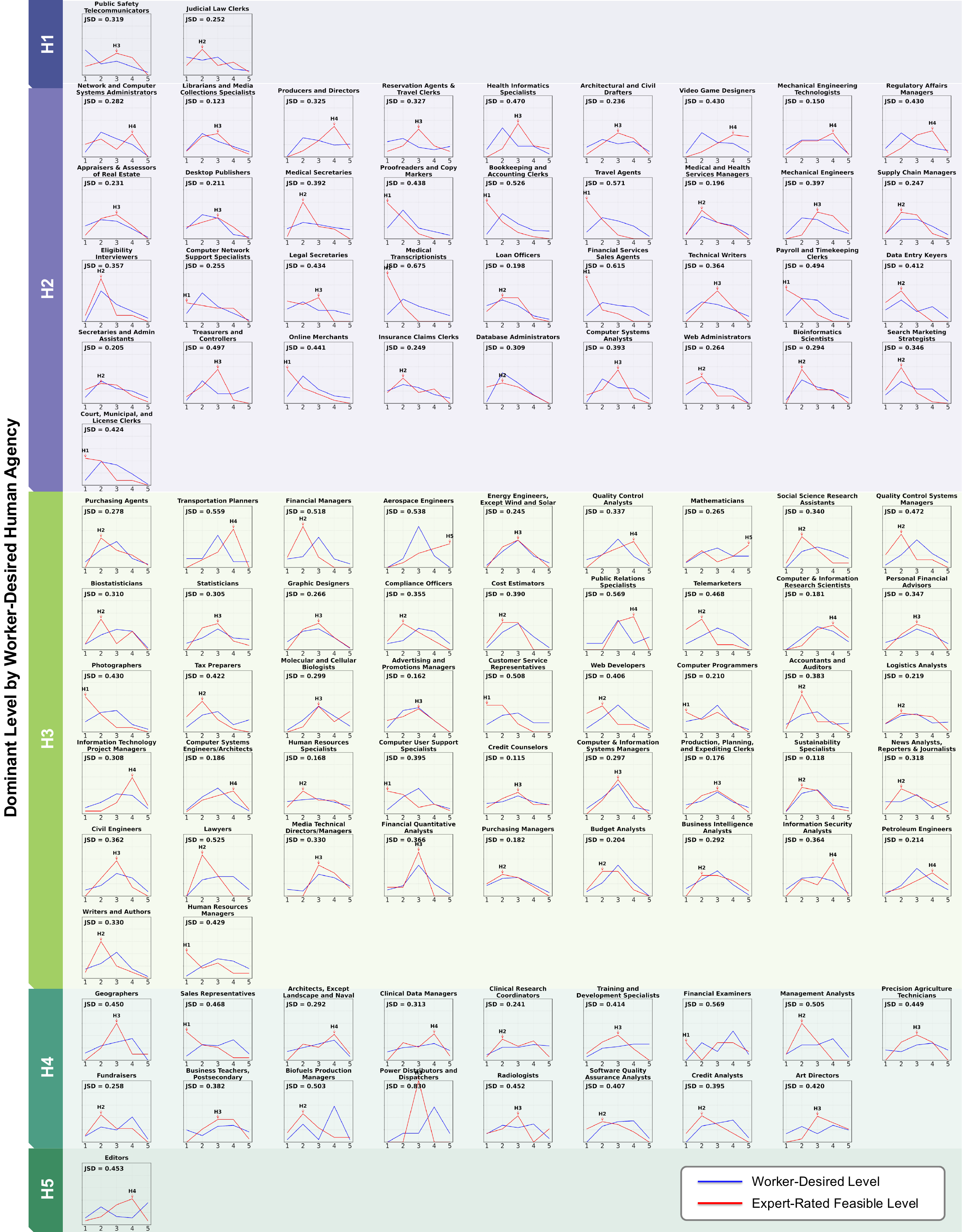}
  \end{minipage}
  \caption{%
    \small{
    \textbf{Distributions of Human Agency Scale (HAS) levels.}
    The Jensen–Shannon Divergence (JSD) quantifies the divergence between the distribution of worker-desired HAS levels ($H_w(t)$) and expert-assessed feasible HAS levels ($H_e(t)$).
  }}
  \label{Fig:has_full_results}
\end{figure*}
\restoregeometry
\clearpage

\subsection{Top 10 Occupations By Worker-Expert Discrepancies in HAS Ratings}

\begin{table*}[h]
\centering
\resizebox{\textwidth}{!}{%
\input{tables/jsd_comparison}
}
\caption{Top 10 occupations with the largest discrepancies between the worker-desired Human Agency Scale levels $H_w(t)$ and AI expert-assessed feasible levels $H_e(t)$. The discrepancy is measured by the Jensen–Shannon divergence (JSD) between these two distributions, where distributions are computed based on all annotated tasks within each occupation.}
\label{table:jsd_comparison}
\end{table*}

\begin{table*}[h]
\centering
\resizebox{\textwidth}{!}{%
\input{tables/skill-info}
}
\caption{Full descriptions of skills covered by the \name database.%
}
\label{table:skill_info}
\end{table*}

\subsection{Task to Skill Mapping}
\label{appendix:skill}
We use the O*NET database to map tasks to their related skills (Generalized Work Activities). In order to create broad groups of common skills across multiple tasks, O*NET defines three levels of work activities: Generalized (GWA), Intermediate (IWA), and Detailed Work Activities (DWA). Each task in the O*NET database is mapped to one or more DWA's, which then corresponds to exactly one GWA. 

For example, the task ``Compile financial data to prepare quarterly budget reports'' is mapped to the DWA ``Prepare financial documents, reports, or budgets.'' This DWA is then associated with the skill (GWA): ``Documenting/Recording Information''. 
Using this mapping, we are able to directly match tasks to their corresponding skills.

\reftab{table:skill_info} provides the full descriptions of the skills shown in \reffig{Fig:skill_shift}, ordered by decreasing required human agency~\citep{tsacoumis2010net}. We exclude physical skills (\ie, ``Identifying Objects, Actions, and Events,'' ``Performing General Physical Activities,'' ``Controlling Machines and Processes,'' ``Inspecting Equipment, Structures, or Materials,'' ``Handling and Moving Objects,'' ``Repairing and Maintaining Mechanical Equipment'') and ``Working with Computers,'' as the audit focuses on tasks that are likely to be exposed to digital AI agents.

\section{Audio Response Analysis}

\subsection{LLM-based Topic Modeling}
\label{sec:topic_modeling}

\begin{table*}[t]
\centering
\input{tables/transcript_analysis_fear}
\caption{Identified concepts with the seed prompt ``The top most common fears that workers have about AI automation in their work''.}
\label{table:transcript_analysis_fear}
\end{table*}

\begin{table*}[t]
\centering
\input{tables/transcript_analysis_collab}
\caption{Identified concepts with the seed prompt ``If workers want AI to help, what type of imagined partnership with AI do they prefer''.}
\label{table:transcript_analysis_collab}
\end{table*}

To analyze our audio transcripts collected from workers, we use LLooM~\citep{lam2024concept}, an LLM-based topic modeling tool that takes unstructured text to extract high-level concepts.

Using LlooM, we first apply a distillation step with \texttt{Claude 3 Opus} on the audio transcripts. This step filters the text to retain only the most relevant quotes based on a seed prompt and performs a summarization of the filtered text. Subsequently, we cluster these summarized quotes using the same LLM to form groups of related concepts. We manually inspect the clusters to ensure the concepts are distinct and merge any overlapping concepts as necessary. \reftab{table:transcript_analysis_fear} and \reftab{table:transcript_analysis_collab} present the top concept groups identified for two seed prompts, corresponding to the analyses in \refsec{sec:automation_desire} and \refsec{sec:has_result}, respectively.

\subsection{Audio Response Examples}
\label{appendix:audio_response}

\subsubsection{Full Transcripts for Direct Quotes in \refsec{sec:automation_desire}}

\input{tables/transcript_for_direct_quote}

\subsubsection{Transcripts from Occupations Exhibiting High Human Agency (HAS) Levels}

As discussed in~\refsec{sec:has_result}, the Human Agency Scale (HAS) spectrum reveals that very few occupations are characterized by a dominant HAS Level 5 (indicating essential human involvement). ``Editors'' is the only occupation where workers predominantly desire H5. According to AI expert assessments, only ``Mathematicians'' and ``Aerospace Engineers'' fall into this category. Below are representative transcripts for these occupations.

\input{tables/transcript_for_high_has_occupation}

%% file: tables/mixedlm_summary.tex
\begin{tabular}{lrrrr|lrrrr}
\toprule
\textbf{Variable} & \textbf{Coef.} & \textbf{Std.Err.} & \textbf{z} & \textbf{P>|z|} & \textbf{Variable} & \textbf{Coef.} & \textbf{Std.Err.} & \textbf{z} & \textbf{P>|z|} \\
\midrule
Intercept & 2.736 & 0.224 & 12.208 & 0.000 & llm\_usage\_by\_type\_\_edit[T.Monthly] & 0.023 & 0.065 & 0.352 & 0.725 \\
gender[T.Male] & 0.070 & 0.037 & 1.909 & 0.056 & llm\_usage\_by\_type\_\_edit[T.Never] & -0.185 & 0.076 & -2.447 & 0.014 \\
gender[T.Other] & -0.210 & 0.356 & -0.591 & 0.555 & llm\_usage\_by\_type\_\_edit[T.Weekly] & -0.094 & 0.047 & -2.009 & 0.045 \\
gender[T.Prefer not to say] & 0.153 & 0.141 & 1.084 & 0.278 & llm\_usage\_by\_type\_\_idea\_generation[T.Monthly] & -0.096 & 0.062 & -1.545 & 0.122 \\
education[T.Bachelor's Degree] & 0.022 & 0.081 & 0.268 & 0.789 & llm\_usage\_by\_type\_\_idea\_generation[T.Never] & 0.072 & 0.076 & 0.952 & 0.341 \\
education[T.Doctorate (e.g., PhD)] & 0.236 & 0.113 & 2.086 & 0.037 & llm\_usage\_by\_type\_\_idea\_generation[T.Weekly] & -0.014 & 0.049 & -0.279 & 0.781 \\
education[T.High School] & 0.037 & 0.117 & 0.319 & 0.750 & llm\_usage\_by\_type\_\_communication[T.Monthly] & -0.042 & 0.066 & -0.641 & 0.521 \\
education[T.Master's Degree] & 0.121 & 0.084 & 1.436 & 0.151 & llm\_usage\_by\_type\_\_communication[T.Never] & -0.052 & 0.067 & -0.785 & 0.432 \\
education[T.Prefer not to say] & 0.157 & 0.221 & 0.711 & 0.477 & llm\_usage\_by\_type\_\_communication[T.Weekly] & -0.051 & 0.047 & -1.084 & 0.278 \\
education[T.Professional Degree (e.g., MD, JD)] & -0.087 & 0.130 & -0.670 & 0.503 & llm\_usage\_by\_type\_\_analysis[T.Monthly] & 0.078 & 0.067 & 1.158 & 0.247 \\
education[T.Some College, No Degree] & -0.125 & 0.093 & -1.340 & 0.180 & llm\_usage\_by\_type\_\_analysis[T.Never] & -0.013 & 0.080 & -0.165 & 0.869 \\
experience[T.3-5 years] & 0.100 & 0.060 & 1.662 & 0.097 & llm\_usage\_by\_type\_\_analysis[T.Weekly] & 0.126 & 0.052 & 2.415 & 0.016 \\
experience[T.6-10 years] & 0.141 & 0.065 & 2.180 & 0.029 & llm\_usage\_by\_type\_\_decision[T.Monthly] & 0.022 & 0.066 & 0.329 & 0.742 \\
experience[T.Less than 1 year] & -0.073 & 0.120 & -0.610 & 0.542 & llm\_usage\_by\_type\_\_decision[T.Never] & -0.085 & 0.077 & -1.107 & 0.268 \\
experience[T.More than 10 years] & 0.229 & 0.067 & 3.415 & 0.001 & llm\_usage\_by\_type\_\_decision[T.Weekly] & -0.122 & 0.054 & -2.268 & 0.023 \\
llm\_familiarity[T.I have some experience using them.] & -0.266 & 0.154 & -1.726 & 0.084 & llm\_usage\_by\_type\_\_coding[T.Monthly] & -0.021 & 0.064 & -0.328 & 0.743 \\
llm\_familiarity[T.I use them regularly.] & -0.266 & 0.156 & -1.711 & 0.087 & llm\_usage\_by\_type\_\_coding[T.Never] & 0.057 & 0.067 & 0.850 & 0.396 \\
llm\_familiarity[T.No, I've never heard of them.] & 1.370 & 0.611 & 2.241 & 0.025 & llm\_usage\_by\_type\_\_coding[T.Weekly] & -0.130 & 0.059 & -2.228 & 0.026 \\
llm\_use\_in\_work[T.Yes, I use them every day in my work.] & 0.077 & 0.053 & 1.457 & 0.145 & llm\_usage\_by\_type\_\_system\_design[T.Monthly] & -0.171 & 0.073 & -2.334 & 0.020 \\
llm\_use\_in\_work[T.Yes, I use them every week in my work.] & 0.025 & 0.051 & 0.493 & 0.622 & llm\_usage\_by\_type\_\_system\_design[T.Never] & -0.173 & 0.073 & -2.362 & 0.018 \\
race[T.Black] & -0.077 & 0.069 & -1.127 & 0.260 & llm\_usage\_by\_type\_\_system\_design[T.Weekly] & -0.131 & 0.070 & -1.858 & 0.063 \\
race[T.Hispanic] & -0.024 & 0.093 & -0.260 & 0.795 & llm\_usage\_by\_type\_\_data\_processing[T.Monthly] & -0.026 & 0.060 & -0.430 & 0.667 \\
race[T.Native American] & -0.209 & 0.166 & -1.257 & 0.209 & llm\_usage\_by\_type\_\_data\_processing[T.Never] & 0.099 & 0.061 & 1.633 & 0.102 \\
race[T.Other] & 0.233 & 0.088 & 2.657 & 0.008 & llm\_usage\_by\_type\_\_data\_processing[T.Weekly] & 0.026 & 0.052 & 0.494 & 0.622 \\
race[T.White] & -0.139 & 0.062 & -2.226 & 0.026 & ai\_tedious\_work\_attitude[T.Somewhat agree] & 0.385 & 0.096 & 3.991 & 0.000 \\
income[T.165K-209K] & 0.209 & 0.080 & 2.621 & 0.009 & ai\_tedious\_work\_attitude[T.Somewhat disagree] & 0.538 & 0.128 & 4.206 & 0.000 \\
income[T.209K-529K] & 0.212 & 0.102 & 2.076 & 0.038 & ai\_tedious\_work\_attitude[T.Strongly agree] & 0.685 & 0.097 & 7.098 & 0.000 \\
income[T.30-60K] & 0.275 & 0.063 & 4.332 & 0.000 & ai\_tedious\_work\_attitude[T.Strongly disagree] & 0.439 & 0.126 & 3.482 & 0.000 \\
income[T.529K+] & 0.687 & 0.181 & 3.792 & 0.000 & ai\_job\_importance\_attitude[T.Somewhat agree] & -0.027 & 0.050 & -0.534 & 0.593 \\
income[T.60-86K] & 0.113 & 0.066 & 1.711 & 0.087 & ai\_job\_importance\_attitude[T.Somewhat disagree] & -0.040 & 0.056 & -0.720 & 0.472 \\
income[T.86K-165K] & 0.089 & 0.064 & 1.403 & 0.161 & ai\_job\_importance\_attitude[T.Strongly agree] & -0.087 & 0.065 & -1.351 & 0.177 \\
income[T.Prefer not to say] & -0.035 & 0.107 & -0.329 & 0.742 & ai\_job\_importance\_attitude[T.Strongly disagree] & -0.012 & 0.071 & -0.172 & 0.864 \\
political\_affiliation[T.Green Party] & -0.198 & 1.157 & -0.171 & 0.864 & ai\_daily\_interest\_attitude[T.Somewhat agree] & 0.128 & 0.074 & 1.724 & 0.085 \\
political\_affiliation[T.Independent] & -0.074 & 0.053 & -1.403 & 0.161 & ai\_daily\_interest\_attitude[T.Somewhat disagree] & 0.027 & 0.103 & 0.262 & 0.793 \\
political\_affiliation[T.Libertarian] & 0.193 & 0.125 & 1.539 & 0.124 & ai\_daily\_interest\_attitude[T.Strongly agree] & 0.415 & 0.078 & 5.308 & 0.000 \\
political\_affiliation[T.No political affliation] & 0.063 & 0.057 & 1.092 & 0.275 & ai\_daily\_interest\_attitude[T.Strongly disagree] & 0.069 & 0.109 & 0.633 & 0.526 \\
political\_affiliation[T.Other] & 0.208 & 0.157 & 1.328 & 0.184 & ai\_suffering\_attitude[T.Somewhat agree] & -0.299 & 0.058 & -5.191 & 0.000 \\
political\_affiliation[T.Prefer not to answer] & 0.136 & 0.070 & 1.942 & 0.052 & ai\_suffering\_attitude[T.Somewhat disagree] & -0.047 & 0.055 & -0.852 & 0.394 \\
political\_affiliation[T.Republican] & 0.054 & 0.048 & 1.140 & 0.254 & ai\_suffering\_attitude[T.Strongly agree] & -0.438 & 0.074 & -5.928 & 0.000 \\
llm\_usage\_by\_type\_\_information\_access[T.Monthly] & -0.082 & 0.065 & -1.256 & 0.209 & ai\_suffering\_attitude[T.Strongly disagree] & -0.023 & 0.059 & -0.391 & 0.696 \\
llm\_usage\_by\_type\_\_information\_access[T.Never] & -0.125 & 0.077 & -1.624 & 0.104 & age & -0.003 & 0.002 & -2.033 & 0.042 \\
llm\_usage\_by\_type\_\_information\_access[T.Weekly] & 0.098 & 0.046 & 2.149 & 0.032 &  &  &  &  &  \\
\bottomrule
\end{tabular}

%% file: prompts/occupation_use_computer.tex
{
\begin{tcolorbox}[colframe=black, colback={rgb,255:red,250;green,250;blue,250}, colbacktitle=gray!50!black, fontupper=\small, fonttitle=\bfseries\color{white}, title=Prompt for Filtering Occupations]
Does this job mainly involve using computers?\\
Answer format: ``Yes'' or ``No''\\
Occupation: \{occupation\}

\end{tcolorbox}%

\captionsetup{type=figure}
\label{prompt:filter_occupation}
}

%% file: prompts/workflow_use_computer.tex
{
\begin{tcolorbox}[colframe=black, colback={rgb,255:red,250;green,250;blue,250}, colbacktitle=gray!50!black, fontupper=\small, fonttitle=\bfseries\color{white}, title=Prompt for Filtering Occupational Tasks]
For \{occupation\}, is it possible to finish its work-related task on a computer?\\
Answer format: ``Yes'' or ``No''\\
Task: \{task\}

\end{tcolorbox}%

\captionsetup{type=figure}
\label{prompt:filter_workflow}
}

%% file: prompts/workflow_mapping.tex
\vspace{4em}
\hypertarget{yc-classifier}{}
\begin{tcolorbox}[colframe=black, colback={rgb,255:red,250;green,250;blue,250}, colbacktitle=gray!50!black, fontupper=\small, fonttitle=\bfseries\color{white}, title=Prompt for YC Company Classifier]
You will be presented with a company description. Your job is to classify if the company is an AI related company or not.\\
An AI-related company is defined as a company that is involved in the research, development, or application of AI.\\
Output a boolean value.\\
------\\
The description of the company: \{company\_description\}\\
The boolean value indicating if the company is an AI-related company:
\end{tcolorbox}

\vspace{4em}

\hypertarget{paper-classifier}{}
\begin{tcolorbox}[colframe=black, colback={rgb,255:red,250;green,250;blue,250}, colbacktitle=gray!50!black, fontupper=\small, fonttitle=\bfseries\color{white}, title=Prompt for AI Agent Paper Classifier]
You will see a paper TITLE and ABSTRACT. Return True if the paper's main contribution is an LLM-driven AGENT SYSTEM, else False.\\
\\
Agent-system criteria (must ALL hold):\\
1. Beyond single turn raw text completion - The LLM's output decides what action or module happens next (planner/controller role), beyond single turn raw text completion.\\
2. Implemented pipeline - A complete system is actually built and run, not merely proposed.\\
3. Task-level evaluation - The paper reports results on the entire system performing its task (automatic metrics or user studies).\\
4. Realistic task - The task matches a plausible real-world workflow or a credible simulated environment.\\
\\
If ANY criterion is missing, output "False".\\
------\\
The title of the paper: \{paper\_title\}\\
The abstract of the paper: \{paper\_abstract\}\\
Whether the paper is an LLM-driven agent system:
\end{tcolorbox}

\newpage
\vspace*{\fill} %

\hypertarget{paper-task-extractor}{}
\begin{tcolorbox}[colframe=black, colback={rgb,255:red,250;green,250;blue,250}, colbacktitle=gray!50!black, fontupper=\small, fonttitle=\bfseries\color{white}, title=Prompt for Paper Task Extractor]
You will be given the TITLE and ABSTRACT of a research paper describing an agent system.  Extract the core task the paper addresses and express it in one concise sentence that begins exactly with: "The paper proposes an agent system to solve the task of...". Your output should be only that sentence, capturing the primary objective of the system.\\
------\\
The title of the paper: \{paper\_title\}\\
The abstract of the paper: \{paper\_abstract\}\\
The core task the paper addresses, expressed in one concise sentence:
\end{tcolorbox}

\vspace{4em}

\hypertarget{company-to-workflow-classifier}{}
\begin{tcolorbox}[colframe=black, colback={rgb,255:red,250;green,250;blue,250}, colbacktitle=gray!50!black, fontupper=\small, fonttitle=\bfseries\color{white}, title=Prompt for Company-to-Task Classifier]
You will receive a brief description of a company and its product/service and an indexed list of workflows. For each workflow, decide whether workers involved in that workflow are a primary or explicitly intended user group of the company's offering. If the link is merely incidental, indirect, or speculative, mark it False. When in doubt, default to False.\\
------\\
The description of the company: \{company\_description\}\\
The list of workflows and their descriptions: \{workflows\}\\
The dictionary of occupations and whether they are the target users of the company:
\end{tcolorbox}

\vspace{4em}

\hypertarget{paper-to-workflow-classifier}{}
\begin{tcolorbox}[colframe=black, colback={rgb,255:red,250;green,250;blue,250}, colbacktitle=gray!50!black, fontupper=\small, fonttitle=\bfseries\color{white}, title=Prompt for Paper-to-Task Classifier]
You will receive a brief description of a task proposed in the paper and an indexed list of workflows. For each workflow, decide whether the task is related to the workflow and the research is applicable to the workflow. If the link is merely incidental, indirect, or speculative, mark it False. When in doubt, default to False.\\
------\\
The description of the task proposed in the paper: \{task\_description\}\\
The list of workflows and their descriptions: \{workflows\}\\
The dictionary of workflows and whether they are related to the task:
\end{tcolorbox}

\vspace*{\fill}

%% file: tables/occupation-list.tex
\begin{tabular}{lclc} 
\toprule
\multicolumn{1}{c}{\textbf{Occupation (O*NET-SOC Title)}}                             & $N$ & \multicolumn{1}{c}{\textbf{Occupation (O*NET-SOC Title)}}      & $N$  \\ 
\midrule
Customer Service Representatives                                                      & 53  & Biostatisticians                                               & 12   \\
Sales Representatives, Wholesale and Manufacturing, Technical and Scientific Products & 35  & Quality Control Analysts                                       & 12   \\
Accountants and Auditors                                                              & 31  & Advertising and Promotions Managers                            & 12   \\
Clinical Research Coordinators                                                        & 30  & Budget Analysts                                                & 11   \\
Medical and Health Services Managers                                                  & 30  & Public Relations Specialists                                   & 11   \\
Computer Programmers                                                                  & 28  & Financial Managers                                             & 11   \\
Web Developers                                                                        & 27  & Logistics Analysts                                             & 11   \\
Computer and Information Systems Managers                                             & 27  & Medical Transcriptionists                                      & 11   \\
Computer Systems Analysts                                                             & 25  & Radiologists                                                   & 11   \\
Purchasing Managers                                                                   & 24  & Eligibility Interviewers, Government Programs                  & 11   \\
Business Teachers, Postsecondary                                                      & 24  & Court, Municipal, and License Clerks                           & 11   \\
Information Technology Project Managers                                               & 24  & Securities, Commodities, and Financial Services Sales Agents   & 11   \\
Financial Quantitative Analysts                                                       & 23  & Sustainability Specialists                                     & 11   \\
Insurance Claims and Policy Processing Clerks                                         & 21  & Web Administrators                                             & 11   \\
Computer and Information Research Scientists                                          & 20  & Geographers                                                    & 11   \\
Legal Secretaries and Administrative Assistants                                       & 19  & Management Analysts                                            & 11   \\
Secretaries and Administrative Assistants, Except Legal, Medical, and Executive       & 18  & Bioinformatics Scientists                                      & 11   \\
Human Resources Specialists                                                           & 18  & News Analysts, Reporters, and Journalists                      & 11   \\
Human Resources Managers                                                              & 17  & Compliance Officers                                            & 11   \\
Business Intelligence Analysts                                                        & 17  & Video Game Designers                                           & 11   \\
Purchasing Agents, Except Wholesale, Retail, and Farm Products                        & 17  & Lawyers                                                        & 11   \\
Statisticians                                                                         & 16  & Proofreaders and Copy Markers                                  & 10   \\
Personal Financial Advisors                                                           & 16  & Architects, Except Landscape and Naval                         & 10   \\
Production, Planning, and Expediting Clerks                                           & 16  & Art Directors                                                  & 10   \\
Computer User Support Specialists                                                     & 16  & Data Entry Keyers                                              & 10   \\
Search Marketing Strategists                                                          & 16  & Public Safety Telecommunicators                                & 10   \\
Architectural and Civil Drafters                                                      & 15  & Producers and Directors                                        & 10   \\
Media Technical Directors/Managers                                                    & 15  & Precision Agriculture Technicians                              & 10   \\
Editors                                                                               & 15  & Credit Counselors                                              & 10   \\
Training and Development Specialists                                                  & 15  & Tax Preparers                                                  & 10   \\
Transportation Planners                                                               & 15  & Health Informatics Specialists                                 & 10   \\
Appraisers and Assessors of Real Estate                                               & 14  & Aerospace Engineers                                            & 10   \\
Fundraisers                                                                           & 14  & Medical Secretaries and Administrative Assistants              & 10   \\
Credit Analysts                                                                       & 14  & Technical Writers                                              & 10   \\
Writers and Authors                                                                   & 14  & Reservation and Transportation Ticket Agents and Travel Clerks & 10   \\
Graphic Designers                                                                     & 14  & Petroleum Engineers                                            & 10   \\
Civil Engineers                                                                       & 14  & Molecular and Cellular Biologists                              & 10   \\
Information Security Analysts                                                         & 14  & Social Science Research Assistants                             & 10   \\
Online Merchants                                                                      & 14  & Financial Examiners                                            & 10   \\
Mathematicians                                                                        & 13  & Telemarketers                                                  & 10   \\
Travel Agents                                                                         & 13  & Payroll and Timekeeping Clerks                                 & 10   \\
Photographers                                                                         & 13  & Quality Control Systems Managers                               & 10   \\
Software Quality Assurance Analysts and Testers                                       & 13  & Judicial Law Clerks                                            & 10   \\
Bookkeeping, Accounting, and Auditing Clerks                                          & 13  & Database Administrators                                        & 10   \\
Desktop Publishers                                                                    & 12  & Power Distributors and Dispatchers                             & 10   \\
Cost Estimators                                                                       & 12  & Energy Engineers, Except Wind and Solar                        & 10   \\
Regulatory Affairs Managers                                                           & 12  & Network and Computer Systems Administrators                    & 10   \\
Mechanical Engineers                                                                  & 12  & Librarians and Media Collections Specialists                   & 10   \\
Supply Chain Managers                                                                 & 12  & Treasurers and Controllers                                     & 10   \\
Clinical Data Managers                                                                & 12  & Biofuels Production Managers                                   & 10   \\
Computer Systems Engineers/Architects                                                 & 12  & Mechanical Engineering Technologists and Technicians           & 10   \\
Loan Officers                                                                         & 12  & Computer Network Support Specialists                           & 10   \\
\bottomrule
\end{tabular}

%% file: tables/worker_distribution.tex
\begin{tabular}{llc} 
\toprule
\multicolumn{1}{c}{\textbf{Recruitment Source}} & \multicolumn{1}{c}{\textbf{Description}}                                                                                                                                                                                                                                                                                                                                                                                   & \multicolumn{1}{c}{\textbf{\# Participants}}  \\ 
\midrule
Prolific                                        & \begin{tabular}[c]{@{}l@{}}We applied the following prescreen criteria: Employment Status (Full-Time or Part-Time);\\Nationality (United States); Fluent Languages (English); Method of Work (computer or\\mobile device used for $\geq$ 25\% of work). We also used the "Industry" prescreen filter to\\ensure participants could only select occupations matching their Prolific profile industry.\end{tabular} & 1060                                          \\
Upwork                                          & We required Upwork workers to be located in the United States.                                                                                                                                                                                                                                                                                                                                                             & 184                                           \\
Word of Mouth                                   & \begin{tabular}[c]{@{}l@{}}We distributed the survey link to LinkedIn users located in the United States and\\distributed flyers within the United States.\end{tabular}                                                                                                                                                                                                                                                    & 434                                           \\
\bottomrule
\end{tabular}

%% file: tables/expert_distribution.tex
\begin{tabular}{lcl} 
\toprule
\multicolumn{1}{c}{\textbf{Career Stage}} & \textbf{Count} & \multicolumn{1}{c}{\textbf{Breakdown}}                                                                                                                                                                                                                                                                                                                                                                                                                                                                                                                                                               \\ 
\midrule
PhD Student                               & 25             & \begin{tabular}[c]{@{}l@{}}1st year (3), 2nd year (2), 3rd year (7), 4th year (6), 5th year or above (7);\\Stanford University (9), Cornell University (2), University of Cambridge (2),\\University of Michigan (2), Carnegie Mellon University (1),\\Northwestern University (1), University of Southern California (1),\\University of Texas at Austin (1), Texas A\&M University (1),\\Purdue University (1), Georgia Institute of Technology (1),\\New York University (1),~University of Illinois Urbana-Champaign (1),\\The Hong Kong University of Science and Technology (1).\end{tabular}  \\
Postdoc                                   & 6              & Stanford University (5), Harvard University (1).                                                                                                                                                                                                                                                                                                                                                                                                                                                                                                                                                     \\
Professor                                 & 6              & \begin{tabular}[c]{@{}l@{}}University of Texas at San Antonio (1), Emory University (1),\\Beihang University (1), University at Buffalo (1),\\Texas A\&M University (1), Massachusetts Institute of Technology (1).\end{tabular}                                                                                                                                                                                                                                                                                                                                                                     \\
Industry Practitioner                     & 15             & \begin{tabular}[c]{@{}l@{}}\textless{} 3 years of experience (8), 3-5 years of experience (5),\\12 years of experience (2);\\Google (4), Amazon (2), Apple (1), Tesla (1), Tencent (1),\\Salesforce (1), xAI (1), Hugging Face (1), Other startups (3)\end{tabular}                                                                                                                                                                                                                                                                                                                                  \\
\bottomrule
\end{tabular}

%% file: tables/top-20-automation-desire.tex
\begin{tabular}{lc} 
\toprule
\multicolumn{1}{c}{\textbf{Task}}                                                                                                                                                                        & \begin{tabular}[c]{@{}c@{}}\textbf{Average}\\\textbf{Automation Desire}\end{tabular}  \\ 
\midrule
Tax Preparers: Schedule appointments with clients.                                                                                                                                                       & 5.00                                                                                  \\ 

\begin{tabular}[c]{@{}l@{}}Public Safety Telecommunicators: Maintain files of information relating to emergency calls, \\such as personnel rosters and emergency call-out and pager files.\end{tabular}  & 4.67                                                                                  \\ 

\begin{tabular}[c]{@{}l@{}}Payroll and Timekeeping Clerks: Issue and record adjustments to pay related to previous \\errors or retroactive increases.\end{tabular}                                       & 4.60                                                                                  \\ 

\begin{tabular}[c]{@{}l@{}}Desktop Publishers: Convert various types of files for printing or for the Internet, using \\computer software\end{tabular}                                                   & 4.50                                                                                  \\ 

Online Merchants: Create or maintain database of customer accounts.                                                                                                                                      & 4.50                                                                                  \\ 

\begin{tabular}[c]{@{}l@{}}Quality Control Systems Managers: Direct the tracking of defects, test results, or other\\regularly reported quality control data.\end{tabular}                               & 4.50                                                                                  \\ 

\begin{tabular}[c]{@{}l@{}}Statisticians: Report results of statistical analyses, including information in the form of \\graphs, charts, and tables.\end{tabular}                                        & 4.50                                                                                  \\ 

\begin{tabular}[c]{@{}l@{}}Computer User Support Specialists: Maintain records of daily data communication transactions, \\problems and remedial actions taken, or installation activities.\end{tabular} & 4.50                                                                                  \\ 

\begin{tabular}[c]{@{}l@{}}Online Merchants: Calculate revenue, sales, and expenses, using financial accounting or \\spreadsheet software.\end{tabular}                                                  & 4.40                                                                                  \\ 

Data Entry Keyers: Store completed documents in appropriate locations.                                                                                                                                   & 4.33                                                                                  \\ 

Petroleum Engineers: Maintain records of drilling and production operations.                                                                                                                             & 4.33                                                                                  \\ 

\begin{tabular}[c]{@{}l@{}}Logistics Analysts: Apply analytic methods or tools to understand, predict, or control logistics \\operations or processes.\end{tabular}                                      & 4.33                                                                                  \\ 

Court, Municipal, and License Clerks: Instruct parties about timing of court appearances.                                                                                                                & 4.33                                                                                  \\ 

Data Entry Keyers: Maintain logs of activities and completed work.                                                                                                                                       & 4.25                                                                                  \\ 

\begin{tabular}[c]{@{}l@{}}Compliance Officers: Prepare correspondence to inform concerned parties of licensing\\decisions or appeals processes.\end{tabular}                                            & 4.25                                                                                  \\ 

\begin{tabular}[c]{@{}l@{}}Web Developers: Back up files from Web sites to local directories for instant recovery in case\\of problems.\end{tabular}                                                     & 4.20                                                                                  \\ 

\begin{tabular}[c]{@{}l@{}}Web Administrators: Back up or modify applications and related data to provide for disaster \\recovery.\end{tabular}                                                          & 4.20                                                                                  \\ 

\begin{tabular}[c]{@{}l@{}}Bioinformatics Scientists: Manipulate publicly accessible, commercial, or proprietary genomic,\\proteomic, or post-genomic databases.\end{tabular}                            & 4.17                                                                                  \\ 

\begin{tabular}[c]{@{}l@{}}Network and Computer Systems Administrators: Perform routine network startup and shutdown\\procedures, and maintain control records.\end{tabular}                             & 4.17                                                                                  \\ 

\begin{tabular}[c]{@{}l@{}}Computer and Information Research Scientists: Approve, prepare, monitor, and adjust \\operational budgets.\end{tabular}                                                       & 4.17                                                                                  \\
\bottomrule
\end{tabular}

%% file: tables/bottom-20-automation-desire.tex
\begin{tabular}{lc} 
\toprule
\multicolumn{1}{c}{\textbf{Task}}                                                                                                                                                                                                                                   & \begin{tabular}[c]{@{}c@{}}\textbf{Average}\\\textbf{Automation Desire}\end{tabular}  \\ 
\midrule
\begin{tabular}[c]{@{}l@{}}Reservation and Transportation Ticket Agents and Travel Clerks: Trace lost, delayed, or \\misdirected baggage for customers.\end{tabular}                                                                                                & 1.50                                                                                  \\
Logistics Analysts: Contact potential vendors to determine material availability.                                                                                                                                                                                   & 1.50                                                                                  \\
Editors: Write text, such as stories, articles, editorials, or newsletters.                                                                                                                                                                                         & 1.60                                                                                  \\
\begin{tabular}[c]{@{}l@{}}Reservation and Transportation Ticket Agents and Travel Clerks: Contact customers or\\travel agents to advise them of travel conveyance changes or to confirm reservations.\end{tabular}                                                 & 1.67                                                                                  \\
\begin{tabular}[c]{@{}l@{}}Video Game Designers: Provide feedback to designers and other colleagues regarding\\game design features.\end{tabular}                                                                                                                   & 1.67                                                                                  \\
\begin{tabular}[c]{@{}l@{}}Librarians and Media Collections Specialists: Code, classify, and catalog books, publications,\\films, audio-visual aids, and other library materials, based on subject matter or standard\\library classification systems.\end{tabular} & 1.67                                                                                  \\
\begin{tabular}[c]{@{}l@{}}Editors: Plan the contents of publications according to the publication's style, editorial\\policy, and publishing requirements.\end{tabular}                                                                                            & 1.67                                                                                  \\
\begin{tabular}[c]{@{}l@{}}Database Administrators: Write and code logical and physical database descriptions and\\specify identifiers of database to management system, or direct others in coding descriptions.\end{tabular}                                      & 1.67                                                                                  \\
\begin{tabular}[c]{@{}l@{}}Graphic Designers: Key information into computer equipment to create layouts for client or\\supervisor.\end{tabular}                                                                                                                     & 1.67                                                                                  \\
\begin{tabular}[c]{@{}l@{}}Mechanical Engineering Technologists and Technicians: Calculate required capacities for\\equipment of proposed system to obtain specified performance and submit data to\\engineering personnel for approval.\end{tabular}               & 1.67                                                                                  \\
\begin{tabular}[c]{@{}l@{}}Secretaries and Administrative Assistants, Except Legal, Medical, and Executive: Establish\\work procedures or schedules and keep track of the daily work of clerical staff.\end{tabular}                                                & 1.67                                                                                  \\
Graphic Designers: Review final layouts and suggest improvements, as needed.                                                                                                                                                                                        & 1.71                                                                                  \\
\begin{tabular}[c]{@{}l@{}}Graphic Designers: Prepare illustrations or rough sketches of material, discussing them with\\clients or supervisors and making necessary changes.\end{tabular}                                                                          & 1.71                                                                                  \\
\begin{tabular}[c]{@{}l@{}}Mechanical Engineering Technologists and Technicians: Interpret engineering sketches,\\specifications, or drawings.\end{tabular}                                                                                                         & 1.75                                                                                  \\
\begin{tabular}[c]{@{}l@{}}Accountants and Auditors: Prepare, examine, or analyze accounting records, financial\\statements, or other financial reports to assess accuracy, completeness, and conformance to\\reporting and procedural standards.\end{tabular}      & 1.75                                                                                  \\
\begin{tabular}[c]{@{}l@{}}Editors: Allocate print space for story text, photos, and illustrations according to space\\parameters and copy significance, using knowledge of layout principles.\end{tabular}                                                         & 1.75                                                                                  \\
\begin{tabular}[c]{@{}l@{}}Producers and Directors: Cut and edit film or tape to integrate component parts into desired\\sequences.\end{tabular}                                                                                                                    & 1.75                                                                                  \\
\begin{tabular}[c]{@{}l@{}}Graphic Designers: Create designs, concepts, and sample layouts, based on knowledge of\\layout principles and esthetic design concepts.\end{tabular}                                                                                     & 1.78                                                                                  \\
\begin{tabular}[c]{@{}l@{}}Librarians and Media Collections Specialists: Locate unusual or unique information in response\\to specific requests.\end{tabular}                                                                                                       & 1.80                                                                                  \\
Editors: Assign topics, events and stories to individual writers or reporters for coverage.                                                                                                                                                                         & 1.80                                                                                  \\
\bottomrule
\end{tabular}

%% file: tables/jsd_comparison.tex
\begin{tabular}{lccc} 
\toprule
\multicolumn{1}{c}{\textbf{Occupation}}                      & \begin{tabular}[c]{@{}c@{}}\textbf{Dominant HAS Level}\\\textbf{By Worker Desire}\end{tabular} & \begin{tabular}[c]{@{}c@{}}\textbf{Dominant HAS Level}\\\textbf{By AI Expert Assessment}\end{tabular} & \textbf{JSD}  \\ 
\midrule
Power Distributors and Dispatchers                           & H4                                                                                          & H3                                                                                                 & 0.830         \\
Medical Transcriptionists                                    & H2                                                                                          & H1                                                                                                 & 0.675         \\
Securities, Commodities, and Financial Services Sales Agents & H2                                                                                          & H1                                                                                                 & 0.615         \\
Travel Agents                                                & H2                                                                                          & H1                                                                                                 & 0.571         \\
Financial Examiners                                          & H4                                                                                          & H1                                                                                                 & 0.569         \\
Public Relations Specialists                                 & H3                                                                                          & H4                                                                                                 & 0.569         \\
Transportation Planners                                      & H3                                                                                          & H4                                                                                                 & 0.559         \\
Aerospace Engineers                                          & H3                                                                                          & H5                                                                                                 & 0.538         \\
Bookkeeping, Accounting, and Auditing Clerks                 & H2                                                                                          & H1                                                                                                 & 0.526         \\
Lawyers                                                      & H3                                                                                          & H2                                                                                                 & 0.525         \\
\bottomrule
\end{tabular}

%% file: tables/skill-info.tex
\begin{tabular}{lccc} 
\toprule
\multicolumn{1}{c}{\textbf{Skill}}                                                                                                                                                                                                                                                                                                               & \begin{tabular}[c]{@{}c@{}}\textbf{\# Supported Tasks}\\\textbf{in WORKBank}\end{tabular} & \begin{tabular}[c]{@{}c@{}}\textbf{Average}\\\textbf{Wage Rank}\end{tabular} & \begin{tabular}[c]{@{}c@{}}\textbf{Average Required }\\\textbf{Human Agency Rank}\end{tabular}  \\ 
\midrule
\begin{tabular}[c]{@{}l@{}}\textbf{Organizing, Planning, and Prioritizing Work}: Developing specific goals and plans to\\prioritize, organize, and accomplish your work.\end{tabular}                                                                                                                                                            & 6                                                                                         & 11                                                                           & 1                                                                                               \\
\begin{tabular}[c]{@{}l@{}}\textbf{Training and Teaching Others}: Identifying the educational needs of others, developing\\formal educational or training programs or classes, and teaching or instructing others.\end{tabular}                                                                                                                  & 4                                                                                         & 21                                                                           & 2                                                                                               \\
\begin{tabular}[c]{@{}l@{}}\textbf{Staffing Organizational Units}: Recruiting, interviewing, selecting, hiring, and\\promoting employees in an organization.\end{tabular}                                                                                                                                                                        & 6                                                                                         & 6                                                                            & 3                                                                                               \\
\begin{tabular}[c]{@{}l@{}}\textbf{Updating and Using Relevant Knowledge}: Keeping up-to-date technically and\\applying new knowledge to your job.\end{tabular}                                                                                                                                                                                  & 19                                                                                        & 2                                                                            & 4                                                                                               \\
\begin{tabular}[c]{@{}l@{}}\textbf{Developing Objectives and Strategies}: Establishing long-range objectives and\\specifying the strategies and actions to achieve them.\end{tabular}                                                                                                                                                            & 17                                                                                        & 3                                                                            & 5                                                                                               \\
\begin{tabular}[c]{@{}l@{}}\textbf{Guiding, Directing, and Motivating Subordinates}: Providing guidance and direction\\to subordinates, including setting performance standards and monitoring performance.\end{tabular}                                                                                                                         & 49                                                                                        & 4                                                                            & 6                                                                                               \\
\begin{tabular}[c]{@{}l@{}}\textbf{Judging the Qualities of Objects, Services, or People}: Assessing the value,\\importance, or quality of things or people.\end{tabular}                                                                                                                                                                        & 47                                                                                        & 5                                                                            & 7                                                                                               \\
\begin{tabular}[c]{@{}l@{}}\textbf{Communicating with Supervisors, Peers, or Subordinates}: Providing information to\\supervisors, coworkers, and subordinates by telephone, in written form, e-mail, or in\\person.\end{tabular}                                                                                                                & 35                                                                                        & 12                                                                           & 8                                                                                               \\
\begin{tabular}[c]{@{}l@{}}\textbf{Providing Consultation and Advice to Others}: Providing guidance and expert advice\\to management or other groups on technical, systems-, or process-related topic\end{tabular}                                                                                                                               & 56                                                                                        & 9                                                                            & 9                                                                                               \\
\begin{tabular}[c]{@{}l@{}}\textbf{Thinking Creatively}: Developing, designing, or creating new applications, ideas,\\relationships, systems, or products, including artistic contributions.\end{tabular}                                                                                                                                        & 117                                                                                       & 7                                                                            & 10                                                                                              \\
\begin{tabular}[c]{@{}l@{}}\textbf{Interpreting the Meaning of Information for Others}: Translating or explaining what\\information means and how it can be used.\end{tabular}                                                                                                                                                                   & 16                                                                                        & 13                                                                           & 11                                                                                              \\
\begin{tabular}[c]{@{}l@{}}\textbf{Making Decisions and Solving Problems}: Analyzing information and evaluating\\results to choose the best solution and solve problems.\end{tabular}                                                                                                                                                            & 52                                                                                        & 10                                                                           & 12                                                                                              \\
\begin{tabular}[c]{@{}l@{}}\textbf{Monitoring Processes, Materials, or Surroundings}: Monitoring and reviewing\\information from materials, events, or the environment, to detect or assess problems.\end{tabular}                                                                                                                               & 40                                                                                        & 8                                                                            & 13                                                                                              \\
\begin{tabular}[c]{@{}l@{}}\textbf{Assisting and Caring for Others}: Providing personal assistance, medical attention,\\emotional support, or other personal care to others such as coworkers, customers, or\\patients.\end{tabular}                                                                                                             & 7                                                                                         & 26                                                                           & 14                                                                                              \\
\begin{tabular}[c]{@{}l@{}}\textbf{Getting Information}: Observing, receiving, and otherwise obtaining information from\\all relevant sources.\end{tabular}                                                                                                                                                                                      & 58                                                                                        & 15                                                                           & 15                                                                                              \\
\begin{tabular}[c]{@{}l@{}}\textbf{Monitoring and Controlling Resources}: Monitoring and controlling resources and\\overseeing the spending of money.\end{tabular}                                                                                                                                                                               & 23                                                                                        & 25                                                                           & 16                                                                                              \\
\begin{tabular}[c]{@{}l@{}}\textbf{Analyzing Data or Information}: Identifying the underlying principles, reasons, or\\facts of information by breaking down information or data into separate parts.\end{tabular}                                                                                                                               & 89                                                                                        & 1                                                                            & 17                                                                                              \\
\begin{tabular}[c]{@{}l@{}}\textbf{Selling or Influencing Others}: Convincing others to buy merchandise/goods or to\\otherwise change their minds or actions.\end{tabular}                                                                                                                                                                       & 8                                                                                         & 16                                                                           & 18                                                                                              \\
\begin{tabular}[c]{@{}l@{}}\textbf{Documenting/Recording Information}: Entering, transcribing, recording, storing, of\\maintaining information in written or electronic/magnetic form.\end{tabular}                                                                                                                                              & 177                                                                                       & 14                                                                           & 19                                                                                              \\
\begin{tabular}[c]{@{}l@{}}\textbf{Evaluating Information to Determine Compliance with Standards}: Using relevant\\information and individual judgment to determine whether events or processes comply\\with laws, regulations, or standards.\end{tabular}                                                                                       & 26                                                                                        & 17                                                                           & 20                                                                                              \\
\begin{tabular}[c]{@{}l@{}}\textbf{Communicating with People Outside the Organization}: Communicating with peo-\\ple outside the organization, representing the organization to customers, the public,\\government, and other external sources. The information can be exchanged in person,\\in writing, or by telephone or e-mail.\end{tabular} & 8                                                                                         & 22                                                                           & 21                                                                                              \\
\begin{tabular}[c]{@{}l@{}}\textbf{Processing Information}: Compiling, coding, categorizing, calculating, tabulating,\\auditing, or verifying information or data.\end{tabular}                                                                                                                                                                  & 41                                                                                        & 20                                                                           & 22                                                                                              \\
\begin{tabular}[c]{@{}l@{}}\textbf{Estimating the Quantifiable Characteristics of Products, Events, or Information}:\\Estimating sizes, distances, and quantities; or determining time, costs, resources, or\\materials needed to perform a work activity.\end{tabular}                                                                          & 14                                                                                        & 23                                                                           & 23                                                                                              \\
\begin{tabular}[c]{@{}l@{}}\textbf{Performing Administrative Activities}: Performing day-to-day administrative tasks\\such as maintaining information files and processing paperwork.\end{tabular}                                                                                                                                               & 24                                                                                        & 18                                                                           & 24                                                                                              \\
\begin{tabular}[c]{@{}l@{}}\textbf{Performing for or Working Directly with the Public}: Performing for people or dealing\\directly with the public. This includes serving customers in restaurants and stores, and\\receiving clients or guests.\end{tabular}                                                                                    & 8                                                                                         & 24                                                                           & 25                                                                                              \\
\begin{tabular}[c]{@{}l@{}}\textbf{Scheduling Work and Activities}: Scheduling events, programs, and activities, as well\\as the work of others.\end{tabular}                                                                                                                                                                                    & 10                                                                                        & 27                                                                           & 26                                                                                              \\
\begin{tabular}[c]{@{}l@{}}\textbf{Establishing and Maintaining Interpersonal Relationships}: Developing constructive\\and cooperative working relationships with others, and maintaining them over time.\end{tabular}                                                                                                                           & 1                                                                                         & 19                                                                           & 27                                                                                              \\
\bottomrule
\end{tabular}

%% file: tables/transcript_analysis_fear.tex
\begin{tabular}{lc} 
\toprule
\multicolumn{1}{c}{\textbf{Concept}}                 & \textbf{Percentage of Summaries}  \\ 
\midrule
Lack of trust in accuracy, reliability or capability & 45.0\%                            \\
Fear of job replacement                              & 23.0\%                            \\
Lack of AI’s human qualities or capabilities         & 16.3\%                            \\
AI not applicable or useful to specific work         & 15.6\%                            \\
\bottomrule
\end{tabular}

%% file: tables/transcript_analysis_collab.tex
\begin{tabular}{lc} 
\toprule
\multicolumn{1}{c}{\textbf{Concept}}      & \textbf{Percentage of Summaries}  \\ 
\midrule
Role-based support                        & 23.1\%                            \\
Assistantship                             & 23.0\%                            \\
Automation                                & 16.5\%                            \\
Separation of Tasks between AI and Humans & 12.0\%                            \\
\bottomrule
\end{tabular}

%% file: tables/transcript_for_direct_quote.tex
\renewcommand{\arraystretch}{1.5} %

\begin{longtable}{p\textwidth} 
\toprule
Art Director with 6–10 Years of Experience: \\
\texttt{\small{So [my work is] basically just first of all looking through all the tasks in the sauna at the start of the day. And then once shoots and, you know, filmings are happening, looking through the footage and pictures, making sure they adhere to brand standards and guidelines and a cohesive voice.}}\\
\texttt{\small{{[}I spend most of my time] Looking through imagery and making sure that it is consistent and always deliverable, and making selects and things of that nature, just establishing a cohesive tone. [I use] So definitely Bridge, you know, Photo Mechanic, Capture One, Photoshop, Asana for the tasks like I mentioned earlier, you know, Gmail, things like that. [For the detailed procedure,] Yeah, looking at the imagery as it's flowing through during the shoot or, you know, filming if it's video, and then from there going through and selecting and culling and, you know, again, only sharing the best imagery that's cohesive.}}\\
\texttt{\small{{[}For AI use,] I don't really, unless it's, you know, in some sort of minor way to help the calling process become easier. \emph{I don't want it to be used for content creation. If anything, I want it to be used for seamlessly maximizing workflow and, you know, making things less repetitive and tedious and arduous with workflow. No content creation.}}} \\
\midrule
Art Director with 3-5 Years of Experience:\\
\texttt{\small{I manage some anime art projects as part of a company's public relations and community strategy for youth engagement. So I work with artists directly, manage projects and merchandise and tabling at events and all that fun stuff.}}\\
\texttt{\small{I do a lot of internal meetings just to make sure everyone's on the same page. It takes up a lot of my time. I also have to scope out projects, find artist to work with especially those we found on social media, figure out how to get in touch and work with them, work with community groups as well, do this type of stuff. And then I also help, well not directly do, but help assist in getting merchandise produced, including preparing artwork for like production and stuff like that.}}\\
\texttt{\small{I spend a lot of time in Microsoft Word, Microsoft Teams, Outlook, which, you know, my company uses Microsoft Office for everything. But then I also use, to communicate with artists, I use the apps that they use. So sometimes that's LINE, sometimes that's Discord, sometimes that's Twitter. And then I personally have my own notion for project management as well.}}\\
\texttt{\small{I spend a lot of time sculpting out projects, so I generally start brainstorming or collecting all my research, gather all my information into Notion first, and then put it into a Word document. The Word document's a little more formal, but I also like make sure that's still approachable to artists, and then I, you know, export that as a PDF. For communicating with artists, that depends on what they use, but most of them use Discord, so it's just back and forth. They send us a sketch, I send that off to my people for, like, feedback very quickly, and then I, you know, get back to them. I sometimes do have my own autonomy to, like, do the final say on what does work and what doesn't work. And then for at least getting everyone on the same page, I spend a lot of time Microsoft Teams. I obviously have to gather some meeting notes, like, write down some job stuff I want to talk about beforehand, make sure there's no surprises to people, it's just communicating and providing regular updates.}}\\
\texttt{\small{\emph{I would never use AI to like replace artists. I would be more for personal [project management] use, if anything}, it's to summarize my tasks, for example, or things like improving my writing, using Apple's writing tools, where I can just revise my writing to be a little more concise, but I would never, let it brainstorm on my behalf, just because I find AI to be very poorly performing on those type of tasks.}}\\

\midrule
Graphic Design with More Than 10 Years of Experience: \\
\texttt{\small{I do basically architecture presentation, like graphic design work, work like typically. It's like layout design, which is organizing content, image or text for AI storyline, diagram or infographics analysis diagram, render enhancement like poster processing, 3D render for a polished look. I also do topography and color scheme using professional fonts and color that align with the project's theme. I also work with board composition, arranging plan, section, elevation and perspectives. I also work digital and print formatting, which is ensuring high quality output for print or physical brands.}}\\
\texttt{\small{So, I basically do architectural presentations. Graphic design work typically includes layout design, diagram, etc. Later enhancement, like post-printing 3D renders for a polished work. I also give topography color scheme both for precision, visibility, and formality. I basically use drawing for autocad and after that for render I use Photoshop and Illustrator and for 3D render I use Lumion.~So basically I spend a lot of time drafting in AutoCAD. The most common tasks just likely include creating floor plans, creating sections, creating elevation, annotation and dimensioning drawings, organizing and managing the layers.}}\\
\texttt{\small{\emph{AI can be a game-changer in data architect workflow, helping to improve efficiency, accuracy and even creativity. But I create my design by myself. For research, I use AI.}}} \\
\bottomrule
\end{longtable}

%% file: tables/transcript_for_high_has_occupation.tex
\renewcommand{\arraystretch}{1.5} %

\begin{longtable}{p\textwidth} 
\toprule
Editor with 3-5 Years of Experience:\\
\texttt{\small{I proofread and copy-edit marketing materials, mostly in the travel and tourism sector. I also do some copywriting and script writing for different ad clients and some light design work.}}\\
\texttt{\small{I look through flyers, brochures, other marketing materials, and I do several passes for mistakes in grammar, in consistency, in flow and clarity. And I make changes on the document, usually a PDF, and send them back to the client. They fix them, they send me another version, and I do several more passes until we've spent enough time and got it perfect. I mainly use Adobe products, PDFs in Adobe Reader. I also use Microsoft Word and some Adobe Suite products, mostly Illustrator and InDesign.}}\\
\texttt{\small{For copy editing, I read through whatever material the client has sent me. I do a pass for basic grammar. I do a pass for clarity and flow, often changing the copy significantly to make it sound better. For proofreading, I go through the materials. Same thing, but with less of a view toward changing the copy and more toward finding errors in grammar and consistency and even design. For copywriting, I make an outline of my ideas for the project and complete a rough draft. Then I spend some time away from it and revise until I have a polished draft for the client.}}\\
\texttt{\small{I'm resistant to using AI in my daily workflow. If I'm forced to use it, I would use it for basic grammar editing, but I would check each suggestion against my own knowledge very carefully and give it full consideration before adopting it as a change.}}\\
\midrule
Editor With More Than 10 Years of Experience:\\
\texttt{\small{So I work in a media company, [masked], and as an editor I make sure that all JavaScript that I'm going to print are formatted correctly, the colors are accurate, and there are no typos.}}\\
\texttt{\small{So a lot of what I do involves sitting at a computer using productivity tools like Adobe Creative Suite, Canva, QuarkXPress, and using the Google Enterprise. So for email, document sharing, I use Google Docs quite a bit for my editing purposes, but I'll also receive files in PDF format. So just working with all the different tools on my computer to get my tasks done every day. So a lot of what I do involves sitting at a computer using productivity tools like Adobe Creative Suite, Canva, QuarkXPress, and using the Google Enterprise. So for email, document sharing, I use Google Docs quite a bit for my editing purposes, but I'll also receive files in PDF format. So just working with all the different tools on my computer to get my tasks done every day.}}\\
\texttt{\small{So one of the most frequent pieces of software I use is Adobe Acrobat, and that is really great for editing PDFs. The next most frequent software I use would be Google Docs. Receiving files through Google Docs, that's a great way to be able to provide updates and edits to annotate the files. And then I would say other tools like Adobe InDesign, Adobe Photoshop, QuarkXPress, Microsoft Publisher. Those are occasionally used, but that's really about all four of the frequently used programs, I would say. So one of the most frequent pieces of software I use is Adobe Acrobat, and that is really great for editing PDFs. The next most frequent software I use would be Google Docs. Receiving files through Google Docs, that's a great way to be able to provide updates and edits to annotate the files. And then I would say other tools like Adobe InDesign, Adobe Photoshop, QuarkXPress, Microsoft Publisher. Those are occasionally used, but that's really about all four of the frequently used programs, I would say.}}\\
\texttt{\small{So, when I'm reviewing PDF documents, I will use the markup tool to add comments and highlight certain sections to make sure that the wording is accurate, or if there's questions regarding the resolution of a photo, I can send that back to mark that up and say, hey, this needs to be a higher resolution photo, it won't print out correctly. So it's just a lot of manual review of every single file before it goes to print to make sure that everything is properly formatted, the colors are accurate, and it will reproduce correctly, just to make sure everything looks good for the customer. So, when I'm reviewing PDF documents, I will use the markup tool to add comments and highlight certain sections to make sure that the wording is accurate, or if there's questions regarding the resolution of a photo, I can send that back to mark that up and say, hey, this needs to be a higher resolution photo, it won't print out correctly. So it's just a lot of manual review of every single file before it goes to print to make sure that everything is properly formatted, the colors are accurate, and it will reproduce correctly, just to make sure everything looks good for the customer.}}\\
\texttt{\small{So, I'm using AI right now when it comes to email, so with the Google suite, there are Google Gemini tools that help with formatting emails. I can take a very simple format for content for email and then using that to expand those topics and make it more of a wordy email. But I would like to be able to use AI more in my proofreading and editing than I am right now, so probably within the next couple months I should be able to do that.}}\\
\midrule
Mathematician With 3-5 Years of Experience:\\
\texttt{\small{I do number theory and algebraic geometry, mostly around long-length programs or categorical long-length programs. [In my daily work, I] read papers and write papers.}}\\
\texttt{\small{Solving a math problem, I don't know [whether there is any specific tool to use], just read papers and have an intuition of what the procedure of philosophy should be and work on it.}}\\
\texttt{\small{To be honest, I think [AI is] useless at this moment.}}\\
\midrule
Mathematician With 3-5 Years of Experience:\\
\texttt{\small{I am studying geometric representation theory and categorical Langlands program. My work involves coming up the problem to work on and learning math tools to help me think of solutions.}}\\
\texttt{\small{I need to spend a lot of time reading papers and learning math tools. I also need to attend the seminar to find collaborators. Then I work on my problem. [In terms of tools,] I mainly use latex. I spend most of my time studying math. Papers in my field can have hundreds of pages - it takes a long time to understand and try to apply the technique.}}\\
\texttt{\small{At present, I don't think AI has any use for mathematicians, at least for DeepSeek and ChatGPT. One core question I am interested in is whether AI can come up with new stuffs that haven't been proposed before rather than solving problems people craft.}}\\
\midrule
Mathematician With 6-10 Years of Experience:\\
\texttt{\small{I used to study number theory, in particular, p-adic Hodge theory in arithmetic geometry. Now, I work on the formalization of p-adic Hodge theory in Lean and also auto-formalization and auto theorem proving.}}\\
\texttt{\small{During formalization, I elaborate, generalize, and fill gaps in mathematical proofs. I design general fomalization frameworks and spend lots of time in Lean coding. Lean coding involves searching theorems, formalizing statements and filling in formalization details in the proof. The last part is the longest part. For auto formalization and formal theorem proving, I spend most of the time coding to establish the LLM's training pipeline and preparing data for the training. I use the interactive theorem prover Lean. I also use LeanSearch and other tools related to Lean to accelerate. I use Python for LLM training and use DeepSeek for coding and debugging.}}\\
\texttt{\small{{[}Here is a concrete example of my workflow:] I formalized a famous number theory definition, called the period rings of Fontaine. I first wrote down a detailed version of the mathematical statements and proofs I need. Splitting the whole formalization project into several smaller goals. For each smaller goal, I generalize and design suitable definitions and lemmas for formalization. Then I begin actual formalization using Lean. I first write down the definitions and state the theorems in Lean without proof. After this, I fill in the proofs backwards, searching the library for existing theorems to use and design patterns to mimic. During formalization, I revise the natural language proof from time to time.}}\\
\texttt{\small{I think a primary AI tool could help me in filling in searching for theorems and design patterns during formalization. A stronger AI tool would do auto-formalization of theorem statements and provide suggestions in designs. An even stronger AI tool would be able to elaborate and fill gaps in mathematical proofs and autoformalize the human proofs. Additionally, an AI tool strong in coding, debugging, and software engineering would help a lot in coding.}}\\
\midrule
Aerospace Engineer With 3-5 Years of Experience:\\
\texttt{\small{I am an aeronautic engineer. I work in the aircraft maintenance, repairs, design aircraft. I work with [masked].}}\\
\texttt{\small{We design aircraft, develop, test, and maintain aircrafts, and the systems that operate within Earth's atmosphere, such as airplane, helicopters, drones, and missiles, though we're not into missiles, though. Our work focuses on making aircraft, machines, very safe and efficient, capable of flight. We use Computer-Aided Design as a tool for Autodex, AutoCAD, Cartier, and Solidwork. And we use Computer Fluid Dynamics. It's ANSYS Fluent, STAR-CNC-MM. What else? We use Finite Element Analysis. It's a tool we use for Nastran.}}\\
\texttt{\small{{[}In my opinion,] AI is going to be very awesome and it's going to make it very easier for us because most of the time, the main problem we have is detecting where the problem is in the engine, you know, so you have to do a lot of manual jobs and all that. So, but if we have AI, you can possibly tell in the dash cam or whatever, you know, you can possibly tell.}}\\
\midrule
Aerospace Engineer With 1-2 Years of Experience:\\
\texttt{\small{I'm a current undergraduate senior and prospective master's student in aerospace engineering, working in guidance, navigation, and controls, so like more simulation, computer programming side of aerospace engineering.}}\\
\texttt{\small{Most of what I do for work has traditionally been programming simulations to evaluate vehicle performance for orbital rockets. And so most of my tasks will be either building out a part of the simulation, programming new features or new testing, or kind of similar types of modeling of different subsystems of the rocket.}}\\
\texttt{\small{In general the tools or software that I use would be Visual Studio Code for the actual programming. The companies I've worked at have used project management tools like JIRA and Confluence. I think also just a lot of internet documentation is useful. And yeah, I very occasionally would use an AI tool like ChatGPT. Generally, my process would be to understand the requirements, which would involve talking to my manager, then kind of going about kind of like pre-reading or other types of information gathering necessary for the task, actual programming, and then like unit testing and other ways of forms of validation for the programming that I completed.}}\\
\texttt{\small{Honestly, I don't use AI too much in my current workflow. I think that the only time that it could come up would be if I'm running into some type of error or bug in my program that I can't find, or kind of a quick piece of code that I could look up how to do, but it's easier to just ask a AI model to generate. But honestly, I use it very, very infrequently.}}\\
\bottomrule
\end{longtable}